\documentclass[12pt]{article}
\usepackage{latexsym}
\usepackage{amsmath,amssymb,amsfonts,amsthm}
\usepackage[english]{babel}
\newcommand{\be}{\begin{equation}}\newcommand{\ee}{\end{equation}}
\newcommand{\bea}{\begin{eqnarray}}\newcommand{\eea}{\end{eqnarray}}
\newcommand{\nn}{\nonumber\\}\newcommand{\p}[1]{(\ref{#1})}
\newcommand{\lb}[1]{\label{#1}}

\def\theequation{\arabic{section}.\arabic{equation}}
\begin{document}

\begin{titlepage}

\vfill

\begin{center}
\baselineskip=16pt {\Large\bf Self-Dual ${\cal N}=2$ Born-Infeld Theory}
\vspace{0.2cm}

{\Large\bf Through Auxiliary Superfields}
\vspace{0.2cm}

\vskip 0.6cm {\large {\sl }} \vskip 10.mm {\bf E.A. Ivanov, $\;$
B.M. Zupnik } \vspace{1cm}

{\it Bogoliubov Laboratory of Theoretical Physics, JINR, \\
141980 Dubna, Moscow Region, Russia\\
}
\vspace{0.3cm}

{\tt eivanov@theor.jinr.ru},  $\;$ {\tt zupnik@theor.jinr.ru}
\end{center}
\vspace{1.6cm}

\par
\begin{center}
{\bf ABSTRACT}
\end{center}
\begin{quote}
There is an evidence that the ${\cal N}=2$ Born-Infeld theory with
spontaneously broken ${\cal N}=4$ supersymmetry exhibits
self-duality. We perform a further check of this hypothesis by
constructing a new representation for the ${\cal N}=2$ Born-Infeld
action through the auxiliary chiral  superfield ${\cal U}$. In such
a formulation, self-duality is equivalent  to $U(1)$ invariance of
the ${\cal U}$ interaction. We explicitly calculate the auxiliary
interaction up to the 10th order and show its $U(1)$ duality
invariance, thus proving that the original action is self-dual to
the same order. We also suggest a new method of recursive
computation of the ${\cal N}=2$ Born-Infeld action in the standard
formulation, based solely on the nonlinear realization of the ${\cal
N}=4$ central charge on the ${\cal N}=2$ superfield strengths ${\cal
W}, \bar{\cal W}$.

\vspace{4.5cm}


\vfill \vfill \vfill \vfill \vfill
\end{quote}
\end{titlepage}

\setcounter{footnote}{0}

\setcounter{page}{1}

\setcounter{equation}0
\section{Introduction}
The ${\cal N}=1$ supersymmetric Born-Infeld (BI) theory \cite{CF,BG,RT} is the notorious  example of theory with the partial spontaneous
breaking of global supersymmetry (PBGS). The relevant superfield action is invariant under the second nonlinearly realized
${\cal N}=1$ supersymmetry and so describes one of the possible patterns of the $d=4$  PBGS ${\cal N}=2 \rightarrow {\cal N}=1\,$,
with the ${\cal N}=1$ spinor gauge superfield strength $W_\alpha$ as the relevant Goldstone fermion. It can be interpreted as the
worldvolume action of the space-filling D3 brane.

It was suggested in \cite{BIK0} that there exists the ${\cal N}=2$ supersymmetric BI action describing the PBGS pattern
${\cal N}=4 \rightarrow {\cal N}=2\,$ and admitting an interpretation as the static-gauge form of the worldvolume action of D3 brane in $D=6$.
The corresponding Goldstone multiplet should be accommodated by the ${\cal N}=2$ Maxwell superfield strength ${\cal W}\,$.

The ${\cal N}=2$ BI action constructed in \cite{Ke} does not reveal
any extra ${\cal N}=2$ supersymmetry and so cannot be regarded as a
candidate for the ${\cal N}=4 \rightarrow {\cal N}=2\,$ BI action.
The group-theoretical setting for the latter (${\cal N}=4, d=4$
superalgebra properly extended by a complex central charge) was
suggested in \cite{BIK00}. In \cite{BIK}, there was proposed the
method of constructing ${\cal N}=2$ BI superfield action within this
approach, such that it is invariant under both the nonlinearly
realized ${\cal N}=4/{\cal N}=2$  supersymmetry and the target space
shift symmetry (symmetry with respect to translations along two
transverse directions of D3 brane). The action was explicitly
restored in a few first orders in the Maxwell   ${\cal N}=2$
superfield strength ${\cal W}\,, \bar{\cal W}$. The terms up to the
8th order were shown to be identical to the analogous recursion
terms found in \cite{KT} from the requirement of ${\cal N}=2$ $U(1)$
self-duality combined with the requirement of the target space shift
invariance. Though the recursive method of \cite{BIK} enables, in
principle, restoring the ${\cal N}=4 \rightarrow {\cal N}=2\,$ BI
action to any order in ${\cal W}$, it remained unclear whether this
action could be given any suggestive closed form.

Recently, there was a revival of interest in the duality-invariant nonlinear extensions of the Maxwell action \cite{GZ,GR,AFZ} and its ${\cal N}=1$
and ${\cal N}=2$ supersymmetric cousins \cite{KT} in connection with the possible crucial role of self-duality in checking the conjectures about ultraviolet
finiteness of ${\cal N}=8$ supergravity and its some lower ${\cal N}$ descendants \cite{BHN,Ren,BN}. We have shown in \cite{IZ3} that the
``nonlinear twisted self-duality constraints'' used in \cite{BN,BCFKR,CarKa} as the systematic method
of constructing self-dual Lagrangians is none other than the equations of motion for auxiliary bispinor fields in the
off-shell formulation of self-duality developed by us in \cite{IZ,IZ2}. A generalization of this
auxiliary-field formulation to ${\cal N}=1, 2$ supersymmetric electrodynamics, with the bispinor fields being promoted
to the chiral spinor or scalar auxiliary superfields,  was recently accomplished in \cite{Ku,ILZ}. The basic advantage of this approach
is that the $U(1)$ duality symmetry\footnote{Or $U(N)$ symmetry - in the case of $N$ Maxwell (super)fields.}
is  realized on the auxiliary (super)fields {\it linearly}, while the full set of self-dual systems is parametrized by $U(1)$ invariant interactions
involving only the auxiliary (super)fields. Another characteristic feature
of this formulation is that many self-dual Lagrangians look much simpler {\it prior} to trading the auxiliary (super)fields for
the Maxwell (super)field  strength. This refers, in particular, to the ${\cal N}=1$ BI action as a typical example of self-dual ${\cal N}=1$
systems.

In application to the ${\cal N}=2$ case, the auxiliary superfield formulation implies that the  superfield action of any self-dual system
can be cast in the following generic form
\bea
{\cal S}({\cal W}, {\cal U}) = {\cal S}_b({\cal W}, {\cal U}) + {\cal I}( {\cal U})\,,  \lb{GenSFN2}
\eea
where ${\cal S}_b$ is some universal bilinear part and ${\cal I}( {\cal U})$ is $U(1)$ invariant interaction encoding
the entire information about the given self-dual system. The standard $({\cal W}, \bar{\cal W})$ form of the action
is reproduced, when eliminating the auxiliary superfield ${\cal U}$ by its equation of motion.

It seems natural to approach the problem of constructing the ``genuine'' ${\cal N}=2$ BI action (with the partially broken ${\cal N}=4$ supersymmetry)
from the duality side, using the auxiliary superfield
formalism as the universal general set-up for the self-dual ${\cal N}=1$ and ${\cal N}=2$ nonlinear electrodynamics actions.
Initiating such a construction is the basic subject of the present paper. We inspect the possibility of putting the ${\cal N}=2$ BI action
into the general self-dual form \p{GenSFN2} and show that the answer is affirmative at least up to the 10th order in the ${\cal W}, \bar{\cal W}$
perturbative expansion of the action.

We start, in section 2, with a brief recalling of what is known about the structure of the ${\cal N}=4/{\cal N}=2$ BI action $S_{BI}({\cal W})$
in the standard ${\cal W}$ representation of refs. \cite{BIK,KT}.
It can be written as a sum of the minimal ${\cal N}=2$ BI action  $S_{\cal X}({\cal W}) = S_2({\cal W}) + I_{\cal X}({\cal W})$ \cite{Ke}
and an additional  nonlinear interaction $\hat{I}({\cal W})$ with the higher-order derivatives,
\bea
S_{BI}({\cal W}) = S_2({\cal W}) + I_{\cal X}({\cal W}) + \hat{I}({\cal W})\,.\lb{N2BIfull}
\eea
The interaction $\hat{I}({\cal W}) $ is an infinite sum of the recursive terms which can
be restored step by step from the requirement of invariance under the second nonlinearly realized ${\cal N}=2$ supersymmetry. In \cite{BIK},
the action $S_{BI}({\cal W})$ was manifestly given up to the 8th order in ${\cal W}, \bar{\cal W}\,$. As a new development, we present the explicit
form of the next, 10th order. We also suggest a new general method of the recursive construction
of the action. It proceeds solely
from the nonlinear realization of the central charge on the superfield strengths ${\cal W}, \bar{\cal W}\,$.

In section 3 we recall the salient features of the $({\cal W}, {\cal U})$ formulation of the self-dual models of ${\cal N}=2$ electrodynamics
and suggest the general form of the hypothetical representation \p{GenSFN2} for the ${\cal N}=2$ BI action:
\bea
{\cal S}_{BI}({\cal W}, {\cal U}) = {\cal S}_b({\cal W}, {\cal U}) + {\cal I}_{BI}({\cal U})\,, \quad
{\cal I}_{BI}({\cal U}) = {\cal I}_{{\cal X}}({\cal U})+
{\cal I}_{\cal R}({\cal U}) + {\cal I}_{\cal Y}({\cal U})\,.\lb{conject}
\eea
Here, ${\cal I}_{{\cal X}}({\cal U})$ is the $({\cal W}, {\cal U})$ ``image'' of the minimal interaction $I_{\cal X}({\cal W})$ and
${\cal I}_{\cal R}({\cal U})$ is obtained through the replacement ${\cal W} \Rightarrow {\cal U}$ in that part of $\hat{I}({\cal W})$ which
is a sum of terms of the highest orders in the $x$-derivatives. This part of $\hat{I}({\cal W})$ can be explicitly written to any order in ${\cal W}$
by the method of ref. \cite{BIK}, as opposed to other parts which involve various descendants of the lower orders in derivatives.
Both ${\cal I}_{{\cal X}}$ and ${\cal I}_{{\cal R}}$ are $U(1)$ invariant and so are guaranteed to give self-dual action after
passing to the $({\cal W}, \bar{\cal W})$ formulation.  The third, unknown interaction part  ${\cal I}_{\cal Y}({\cal U})$
is responsible for some extra possible terms in  $\hat{I}({\cal W})$ in \p{N2BIfull} which cannot be generated
by the previous two $U(1)$ invariant interactions.

In section 4 we cast the action of refs. \cite{Ke} in  the formalism with the auxiliary chiral ${\cal N}=2$
superfields ${\cal U}$. The $({\cal W},{\cal U})$
representation of this action is analogous to
our auxiliary representation of the ${\cal N}=1$ BI action \cite{ILZ}.
We construct, as a series in the auxiliary superfields,  the corresponding ${\cal N}=2$ $U(1)$ invariant interaction ${\cal I}_{{\cal X}}({\cal U})$,
which reproduces the action $S_{\cal X}({\cal W})$ of \cite{Ke} after eliminating the auxiliary superfield ${\cal U}$ by its equation
of motion in the total minimal $({\cal W},{\cal U})$ action ${\cal S}_{{\cal X}}({\cal W}, {\cal U}) = {\cal S}_b({\cal W}, {\cal U}) +
{\cal I}_{\cal X}({\cal U})\,$. We explicitly present ${\cal I}_{{\cal X}}({\cal U})$ up to the 16th order, which,
after going to the conventional ${\cal W}$ action, is capable to reproduce the latter up to the 18th order.

In section 5 we discuss the  $({\cal W}, {\cal U})$ representation
for the full ``genuine'' ${\cal N}=2$ BI action \p{N2BIfull}. We
start with the general form \p{conject} without the unknown
interaction part ${\cal I}_{\cal Y}$. Substitution of the corresponding perturbative solution of the
auxiliary equation for ${\cal U}$ into this action yields, up to the 8th order in ${\cal W}, \bar {\cal W}$,  just the action
$S_{BI}({\cal W})= S_{\cal X}({\cal W}) + \hat{I}({\cal W})\,$, such that all descendants appear in $\hat{I}({\cal W})$
with the correct coefficients.
Unfortunately, starting from the 10th order, we observe a deviation from the genuine ${\cal N}=4/{\cal N}=2$ BI action, which requires
adding the proper auxiliary interaction ${\cal I}_{\cal Y}({\cal U})$.
It is remarkable that such ${\cal I}_{\cal Y}({\cal U})$ indeed exists, and we give explicitly the correction terms
${\cal I}_{\cal Y}^{(10)}({\cal U})$ which prove to be manifestly $U(1)$  invariant. This means that the self-duality of
the ${\cal N}=4/{\cal N}=2$ BI action has been checked up to the 10th order.
In the next orders in ${\cal U}, \bar{\cal U}$ we expect similar correction terms ${\cal I}_{\cal Y}^{(2n)}({\cal U})\,, \,n\geq 6\,,$  too.
At present we are not aware of any systematic way of finding out such auxiliary superfield corrections.

\setcounter{equation}0
\section{${\cal N}=2$ BI theory and spontaneous breaking of ${\cal N}=4$
supersymmetry}

\subsection{The general setting}
The superfield ${\cal N}=2$ BI theory with the second nonlinearly realized ${\cal N}=2$ supersymmetry
was constructed in \cite{BIK},
starting from an infinite-dimensional representation of the full centrally extended ${\cal N}=4$
supersymmetry on chiral superfields \cite{BIK00,BIK}. The  ${\cal N}=4/{\cal N}=2$ transformations
of the chiral superfield strength ${\cal W}$ include the constant shift
of the scalar field (it is associated with the central charge in the ${\cal N}=4$ superalgebra).
The invariance under such a shift was earlier suggested in \cite{KT} as the basic
principle selecting the ${\cal N}=4/{\cal N}=2$ BI action in the plethora of the ${\cal N}=2$ superfield $U(1)$ self-dual actions.
The BI actions derived within these two approaches were found to coincide in a few lowest orders in ${\cal W}, \bar{\cal W}$.

Our conventions for the ${\cal N}=2$ superspace and the ${\cal N}=2$ gauge superfield strengths are described in Appendix A.
We use the free superfield action
\bea
{S}_{2}({\cal W})=\frac14\int d^8{\cal Z}{\cal W}^2
+\frac14\int d^8\bar{\cal Z}\bar{\cal W}^2\,,\lb{free1}
\eea
which yields the correctly normalized component free action.
The full nonlinear action $S_{BI}$ in our notations differs from that of
\cite{BIK} by the factor $1/4$ and the replacement $\Box \rightarrow 2\Box\,$. Note that we ascribe to ${\cal W}$ the non-standard dimension,
$[{\cal W}] = -1$ (in the mass units);
the correct dimension of the action is ensured due to the implicit presence of a dimensionful coupling constant
which, for simplicity,  has been put  equal to 1 hereafter (Appendix A).

We will need both the general and the chiral superspace forms of the BI action
\bea
&&{S}_{BI}({\cal W})= {S}_{2}({\cal W})+ {I}_{BI}({\cal W}) =
\frac14 \int d^8{\cal Z}{\cal A}_0
+\frac14 \int d^8\bar{\cal Z}\bar{\cal A}_0\,, \lb{DeF1} \\
&&{I}_{BI}({\cal W}) =  \int d^{12}Z L_{BI}({\cal W})\,,\lb{defIBI}
\eea
where\footnote{We use the short-hand notations $\bar{D}^4, D^4$ for
the maximal powers of the ${\cal N}=2$ spinor covariant derivatives, see Appendix A.}
\bea
L_{BI}=\sum\limits_{n=2}^\infty L^{(2n)},\quad
{\cal A}_0({\cal W})=\sum\limits_{n=1}^\infty{\cal A}^{(2n)}_0 =
{\cal W}^2+ 2\bar{D}^4L_{BI}\,.\lb{defIBI2}
\eea
The upper index of $L^{(2n)}$ and ${\cal A}_0^{(2n)}$ denotes
the order in ${\cal W}, \bar{\cal W}\,$. The object $L_{BI}$, as defined in \p{defIBI2}, is generically complex, but we will
see that its imaginary part
is a total derivative and so does not contribute to the action $S_{BI}\,$,
$$
{I}_{BI} = \int d^{12}Z\, L_{BI} = \frac12\int d^{12}Z\, (L_{BI}+\bar{L}_{BI}).
$$

The proof of the ${\cal N}=4$ supersymmetry of the action \p{DeF1} is based on the assertion \cite{BIK} that ${\cal W}$ together with ${\cal A}_0$ belong
to an infinite-dimensional linear multiplet of the ${\cal N}=4$ Poincar\'e superalgebra extended by a complex central charge generator;
the latter is assumed to be spontaneously broken (together with the ${\cal N}=4/{\cal N}=2$ part of supersymmetry), so that ${\cal W}$ plays
the role of Goldstone superfield associated with this generator. The ${\cal N}=4/{\cal N}=2$ variations of ${\cal W}$ and ${\cal A}_0$ are given by \cite{BIK}
\bea
&&\delta_f{\cal W}=f(1-\frac12\bar{D}^4\bar{\cal A}_0)+\frac12\bar{f}\Box{\cal A}_0
+\frac{i}4\bar{D}^{\dot\alpha}_k\bar{f}D^{k\alpha}\partial_{\alpha\dot\alpha}
{\cal A}_0,\lb{B1}\\
&&\delta_f{\cal A}_0=2f{\cal W}+\frac12\bar{f}\Box{\cal A}_{1}
+\frac{i}4\bar{D}^{\dot\alpha}_k\bar{f}D^{k\alpha}
\partial_{\alpha\dot\alpha}{\cal A}_{1}\,,\lb{B3}
\eea
where ${\cal A}_1$ is the next chiral superfield component of the linear ${\cal N}=4$ multiplet just mentioned and
\bea
f=c+2i\theta_k^\alpha\xi^k_\alpha\,,\quad \bar{f}=\bar{c}+2i\bar\theta^{k\dot\alpha}\bar\xi_{k\dot\alpha}\,,
\eea
with $c$ and $\xi^k_\alpha\,, k =1,2\,,$ being parameters of the spontaneously broken symmetries. The infinite sequence
of chiral superfields ${\cal A}_n, \;n\geq 1\,,$
have the following transformation laws
\bea
\delta_f{\cal A}_n=2f{\cal A}_{n-1}+\frac12\bar{f}\Box{\cal A}_{n+1}
+\frac{i}4\bar{D}^{\dot\alpha}_k\bar{f}D^{k\alpha}
\partial_{\alpha\dot\alpha}{\cal A}_{n+1}\,.\lb{B4}
\eea
It is straightforward to check that the Lie bracket of two nonlinear supersymmetry transformations of ${\cal W}$ gives the standard $x$ translation
\bea
[\delta_1,\delta_2]{\cal W}=\frac{i}2[(\bar{D}^{\dot\alpha}_k
\bar{f}_2)(D^{k\alpha}f_1)-(\bar{D}^{\dot\alpha}_k
\bar{f}_1)(D^{k\alpha}f_2)]\partial_{\alpha\dot\alpha}{\cal W}=
2i[\xi^{k\alpha}_1\bar{\xi}^{\dot\alpha}_{k2}
-\xi^{k\alpha}_2\bar{\xi}^{\dot\alpha}_{k1}]\partial_{\alpha\dot\alpha}{\cal W}\,,\lb{B2}
\eea
where the identity \p{A10} was used. The transformations of ${\cal A}_n, \;n\geq 0\,,$ have the same closure
\bea
[\delta_1,\delta_2]{\cal A}_n=
2i[\xi^{k\alpha}_1\bar{\xi}^{\dot\alpha}_{k2}
-\xi^{k\alpha}_2\bar{\xi}^{\dot\alpha}_{k1}]\partial_{\alpha\dot\alpha}
{\cal A}_n.
\eea
The action \p{DeF1} is invariant under \p{B3}, taking into account the Bianchi identity \p{A9} and its corollary \p{A10}.

The most difficult step is to express  ${\cal A}_0$ and all subsequent
superfields ${\cal A}_n$ in terms of ${\cal W}, \bar{\cal W}$ and their ordinary and spinor derivatives.
This is achieved by imposing an infinite set of the ${\cal N}=4$ supersymmetric
constraints on ${\cal A}_n$. The first, basic  constraint reads:
\bea
\Phi_0={\cal A}_0-{\cal W}^2-\frac12{\cal A}_0\bar{D}^4\bar{\cal A}_0
-\bar{D}^4\sum\limits_{n=1}\frac{(-1)^n}{ 2^{2n+1}}{\cal A}_n\Box^n
\bar{\cal A}_n=0\,.\lb{B7}
\eea
The higher-order recursion conditions are more complicated, e.g.,
\bea
&& \Phi_1 = \Box{\cal A}_1 +2({\cal A}_0\Box{\cal W} -{\cal W}\Box{\cal A}_0) \nn
&& -\,\bar{D}^4\sum\limits_{n=0}\frac{(-1)^n}{2^{2n+1}}(
\Box{\cal A}_{n+1}\Box^{n}\bar{\cal A}_{n}-{\cal A}_{n+1}\Box^{n+1}\bar{\cal A}_{n})=0\,,\lb{B8} \\
&&\Phi_2=\Box^2{\cal A}_2 +2({\cal A}_0\Box^2{\cal A}_0-\Box{\cal A}_0\Box{\cal A}_0
+2\Box{\cal W}\Box{\cal A}_1-{\cal W}\Box^2{\cal A}_1-{\cal A}_1\Box^2{\cal W})\nn
&&-\,\bar{D}^4\sum\limits_{n=0}\frac{(-1)^n}{2^{2n+1}}(\Box^2{\cal A}_{n+2}\Box^n
\bar{\cal A}_n
-2\Box{\cal A}_{n+2}\Box^{n+1}\bar{\cal A}_n+
{\cal A}_{n+2}\Box^{n+2}\bar{\cal A}_n)=0\,.\lb{B9}
\eea
The next constraints have the generic form $\Phi_n= \Box^n{\cal A}_n + \ldots =0\,$. This infinite set of constraints,
in parallel with \p{B7}, is required by ${\cal N}=4$ supersymmetry, and it allows one to recursively  express ${\cal A}_0$, $\Box {\cal A}_1$ and
$\Box^n{\cal A}_n\,, n \geq 2\,,$ in terms of ${\cal W}, \bar{\cal W}$ and their derivatives.

The perturbative solution for any chiral superfield ${\cal A}_n$ can be written as the following series
\bea
&&{\cal A}_n=\sum\limits_{m=1}^\infty{\cal A}_n^{(n+2m)}\,,\lb{B10}
\eea
where, as before,  the term ${\cal A}_n^{(n+2m)}$ is of order $(n + 2m)$ in ${\cal W}$ and $\bar{\cal W}$.
Substitution of these series into the original set of nonlinear constraints
\p{B7}, \p{B8}, \p{B9},\ldots
gives the double-index chains of recursion relations, in particular,
\bea
&&{\cal A}_0^{(2m)}-
\bar{D}^4\sum\limits_{n=0}^{m-2}\frac{(-1)^n}{ 2^{2n+1}}
\sum\limits_{r=0}^{m-n-2}{\cal A}^{(2m-n-2r-2)}_n\Box^n
\bar{\cal A}^{(n+2r+2)}_n=0\,,\quad m\geq 2\,,\lb{B7b}
\eea
whence, following the definition \p{defIBI2},
\bea
L^{(2m)} = \sum\limits_{n=0}^{m-2}\frac{(-1)^n}{ 2^{2n+2}}
\sum\limits_{r=0}^{m-n-2}{\cal A}^{(2m-n-2r-2)}_n\Box^n
\bar{\cal A}^{(n+2r+2)}_n\,,\quad m\geq 2\,.\lb{B7bb}
\eea
The similar recursions relations can be written for the constraints which
start with $\Box^n{\cal A}_n\,$. {}From the representation \p{B7bb}
it is easy to check that the imaginary parts of $L^{(2m)}$ are indeed total
derivatives and therefore do not contribute to the perturbative expansion of
$I_{BI}({\cal W})$ in \p{defIBI},
\be
\int d^{12}ZL^{(2m)}=\int d^{12}Z\bar{L}^{(2m)}.\lb{ReaL}
\ee

\subsection{Explicit expressions for ${\cal A}_n$}
The characteristic feature of the explicit expressions for terms of different orders in $\Box^n{\cal A}_n({\cal W})$ is that these expressions
can always be represented as $\Box^n$ of something. As a result, the powers of $\Box$ can be taken off from both sides of the relevant equalities,
yielding the explicit expressions for ${\cal A}_n\,$.
The  solutions for the lowest terms in ${\cal A}_n$ for $n\leq 3$ were constructed
in \cite{BIK}. In our conventions, they are
\bea
{\cal A}_0^{(4)}&=& \frac12\bar{D}^4({\cal W}^2\bar{\cal W}^2),\;\;
{\cal A}_0^{(6)}=\frac14\bar{D}^4\left[{\cal W}^2\bar{\cal W}^2(D^4{\cal W}^2
+\bar{D}^4\bar{\cal W}^2)-\frac29{\cal W}^3\Box\bar{\cal W}^3\right],\lb{A046} \\
{\cal A}_0^{(8)} &=& \bar{D}^4\left[
\frac18{\cal W}^2\bar{\cal W}^2(D^4{\cal W}^2)^2
+\frac14{\cal W}^2\bar{\cal W}^2 (D^4{\cal W}^2) (\bar{D}^4\bar{\cal W}^2)
+\frac18{\cal W}^2\bar{\cal W}^2(\bar{D}^4\bar{\cal W}^2)^2
\right.\nn
&&
+\,\frac18{\cal W}^2\bar{\cal W}^2D^4({\cal W}^2 \bar{D}^4\bar{\cal W}^2)
-\frac1{36}{\cal W}^2\bar{\cal W}^3\Box D^4{\cal W}^3\nn
&&
-\,\frac1{18}{\cal W}^3\Box(\bar{\cal W}^3 D^4{\cal W}^2)
-\frac1{12}{\cal W}^3(\bar{D}^4\bar{\cal W}^2)\Box\bar{\cal W}^3
\left.+ \frac1{288}{\cal W}^4\Box^2\bar{\cal W}^4\right], \lb{A08}\\
{\cal A}_1^{(3)}&=&\frac23{\cal W}^3,\quad {\cal A}_1^{(5)}=
\frac23\bar{D}^4({\cal W}^3\bar{\cal W}^2),\nn
{\cal A}_1^{(7)} &=& \bar{D}^4\left[\frac12{\cal W}^3\bar{\cal W}^2
\bar{D}^4\bar{\cal W}^2
+\frac13{\cal W}^3\bar{\cal W}^2 D^4{\cal W}^2
-\frac1{12}{\cal W}^4 \Box\bar{\cal W}^3 \right],\nn
{\cal A}_2^{(4)} &=& \frac13{\cal W}^4,\quad {\cal A}_2^{(6)}=
\frac12\bar{D}^4({\cal W}^4\bar{\cal W}^2),\quad
{\cal A}_3^{(5)}=\frac2{15}{\cal W}^5\,. \lb{A013}
\eea

Knowing these expressions is sufficient for restoring ${\cal A}_0^{(10)}\,$. Following \cite{BIK}, the implicit form of
the latter can be found from the general recursion formula \p{B7b} as
\bea
{\cal A}_0^{(10)} &=& \frac12 \bar D^4 \Big\{{\cal W}^2 \bar{\cal A}_0^{(8)} + \bar{\cal W}^2 {\cal A}_0^{(8)} + {\cal A}_0^{(4)} \bar{\cal A}_0^{(6)} +
{\cal A}_0^{(6)} \bar{\cal A}_0^{(4)} \nn
&& -\,\frac14 [{\cal A}_1^{(3)} \Box \bar{\cal A}_1^{(7)} + {\cal A}_1^{(7)} \Box \bar{\cal A}_1^{(3)} + {\cal A}_1^{(5)} \Box \bar{\cal A}_1^{(5)}] \nn
&& + \frac{1}{16} [{\cal A}_2^{(4)} \Box^2 \bar{\cal A}_2^{(6)} + {\cal A}_2^{(6)} \Box^2 \bar{\cal A}_2^{(4)} ] - \frac{1}{64} {\cal A}_3^{(5)} \Box^3
\bar{\cal A}_3^{(5)} \Big\}. \lb{A100}
\eea

The explicit expression for ${\cal A}_0^{(10)}$ is rather complicated and for this reason was not given in \cite{BIK}. For our further purposes,
it is instructive to present such an expression. It can be written as a sum of three terms
\bea
{\cal A}_0^{(10)} = {\cal X}^{(10)} + {\cal R}^{(10)} + {\cal Y}^{(10)}, \lb{divid}
\eea
with
\bea
{\cal X}^{(10)}&=& \bar{D}^4\Big\{\frac1{16}{\cal W}^2\bar{{\cal W}}^2(D^4{\cal W}^2)^3
+\frac1{8}{\cal W}^2\bar{{\cal W}}^2(\bar{D}^4\bar{{\cal W}}^2)(D^4{\cal W}^2)^2  \nn
&&+\,\frac3{16}{\cal W}^2\bar{{\cal W}}^2 (D^4{\cal W}^2) (\bar{D}^4\bar{{\cal W}}^2)^2
+\frac1{16}{\cal W}^2\bar{{\cal W}}^2(\bar{D}^4\bar{{\cal W}}^2)^3
\nn
&&+\,\frac1{16}{\cal W}^2\bar{{\cal W}}^2D^4
[{\cal W}^2(\bar{D}^4\bar{{\cal W}}^2)^2]
+\frac1{8}{\cal W}^2\bar{{\cal W}}^2 (D^4{\cal W}^2) D^4({\cal W}^2 \bar{D}^4\bar{{\cal W}}^2)
\nn
&&+\,
\frac1{8}{\cal W}^2\bar{{\cal W}}^2 (\bar{D}^4\bar{{\cal W}}^2) D^4({\cal W}^2 \bar{D}^4\bar{{\cal W}}^2)
+\frac1{16}{\cal W}^2\bar{{\cal W}}^2
(D^4{\cal W}^2) \bar{D}^4(\bar{{\cal W}}^2 D^4{\cal W}^2)
\nn
&&
+\,\frac1{16}{\cal W}^2\bar{{\cal W}}^2D^4[{\cal W}^2\bar{D}^4(\bar{{\cal W}}^2D^4{\cal W}^2)]\Big\},  \lb{ket10}\\
{\cal R}^{(10)} &=& \bar D^4\Big\{ -\frac{1}{7200} {\cal W}^5\Box^3\bar{{\cal W} }^5\Big\}\,, \lb{R10} \\
{\cal Y}^{(10)} &=& \bar D^4\Big\{
-\frac1{24}{\cal W}^2\bar{{\cal W}}^3 (D^4{\cal W}^2) D^4\Box {\cal W}^3
-\frac1{24}{\cal W}^3\Box D^4(\bar{{\cal W}}^3{\cal W}^2D^4{\cal W}^2)\nn
&&-\,\frac1{72}{\cal W}^2\bar{{\cal W}}^2D^4({\cal W}^3\Box \bar{D}^4\bar{{\cal W}}^3)
-\frac1{36}{\cal W}^2\bar{{\cal W}}^3D^4 \Box({\cal W}^3\bar{D}^4\bar{{\cal W}}^2)\nn
&&-\,\frac1{72}{\cal W}^2 (\bar{D}^4\bar{{\cal W}}^2) D^4(\bar{{\cal W}}^3\Box {\cal W}^3)
-\frac1{12}{\cal W}^3 (\bar{D}^4\bar{{\cal W}}^2)D^4\Box(\bar{{\cal W}}^3{\cal W}^2)
\nn
&&-\,\frac1{24}{\cal W}^3(\Box\bar{{\cal W}}^3)\bar{D}^4(\bar{{\cal W}}^2D^4{\cal W}^2)
-\frac1{72}{\cal W}^2\bar{{\cal W}}^3 (\bar{D}^4\bar{{\cal W}}^2)\Box D^4{\cal W}^3\nonumber \\
&&-\,\frac1{12}{\cal W}^3 (\bar{D}^4\bar{{\cal W}}^2)^2\Box\bar{{\cal W}}^3
-\frac1{36}{\cal W}^3\Box D^4(\bar{{\cal W}}^3{\cal W}^2 \bar{D}^4\bar{{\cal W}}^2) \nn
&&+\, \frac1{144}{\cal W}^3\Box D^4(\bar{{\cal W}}^4\Box {\cal W}^3)+ \frac1{144}{\cal W}^4(\Box\bar{{\cal W}}^3) \Box\bar{D}^4\bar{{\cal W}}^3 +
\frac1{144}{\cal W}^4 (\bar{D}^4\bar{{\cal W}}^2) \Box^2\bar{{\cal W}}^4\nn
&&+\,\frac1{576}{\cal W}^2\bar{{\cal W}}^4D^4\Box^2{\cal W}^4
+\frac1{192}{\cal W}^4
\Box^2(\bar{{\cal W}}^4 D^4{\cal W}^2)\Big\}. \lb{Y10}
\eea

The basic differences between these three types of terms are as follows. The ${\cal X}$ term contains no box operators inside the curly brackets,
only the operators $D^4$ and $\bar D^4$ are present there; the ${\cal R}$ term contains only box operators; the ${\cal Y}$ term is mixed,
it involves both
the box and the $D^4, \bar D^4$ operators. As is seen from eqs. \p{A046}, \p{A08} a similar division into three such terms is also valid
for ${\cal A}_0^{(4)}, {\cal A}_0^{(6)}$ and ${\cal A}_0^{(8)}\,$. This reflects the general property that the full chiral density  ${\cal A}_0$,
as a consequence of
the constraint \p{B7}, admits the splitting
\bea
{\cal A}_0 = {\cal X} + {\cal R} + {\cal Y}. \lb{splt}
\eea
Here, the superfield ${\cal X}$ is defined by the equation which is a
truncation of the constraint \p{B7}, such that all the terms containing $\Box$
are omitted,
\be
{\cal X} = {\cal W}^2 +\frac12{\cal X} \bar{D}^4\bar{\cal X}\,. \lb{ketov}
\ee
We study eq. \p{ketov} in some detail in section 4. The part ${\cal X}$ also
accounts for the free action ${\cal W}^2$, as well as for the
quartic interaction ${\cal A}_0^{(4)} \sim \bar D^4 ({\cal W}^2\bar{{\cal W}}^2)$.
  The superfield ${\cal R}$ originates from the terms
with ${\cal A}_n, \,n\geq 1\,,$ in \p{B7}:
\be
{\cal R} = 2 \bar{D}^4\sum_{n=3}^\infty (-1)^{n} \frac{1}{(n!)^2} {\cal W}^n
\Box^{n-2}\bar{{\cal W}}^n. \lb{Rgen}
\ee
 The remaining superfield piece ${\cal Y}$ collects, in its perturbative expansion, the mixed terms which
are not combined into any obvious series\footnote{One can still find the series representation for some simple terms in ${\cal Y}$.}.
It contributes to the interaction $I_{BI}$ from the eighth order.

While constructing the auxiliary superfield formulation of the ${\cal N}=4/{\cal N}=2$ BI action in section 5, we will essentially
make use of the general splitting \p{splt}.

As the last topic of this subsection, we will present the explicit form of the BI interaction ${I}_{BI}$ up to the 10th order (with taking into account the
simplifications arising after integrating by parts and grouping similar terms):
\bea
{I}_{BI}^{(4)} &=& \frac14 \int d^{12}Z\, {\cal W}^2 \bar{{\cal W}}^2\,, \lb{4ord} \\
{I}_{BI}^{(6)} &=&
\frac18\int d^{12}Z \Big[{\cal W}^2\bar{{\cal W}}^2(D^4{\cal W}^2 + \bar D^4\bar{{\cal W}}^2) - \frac29{\cal W}^3\Box\bar{{\cal W}}^3\Big], \lb{6ord} \\
{I}_{BI}^{(8)} &=& \frac1{16}\int d^{12}Z \Big\{{\cal W}^2\bar{{\cal W}}^2[(D^4{\cal W}^2)^2 + (\bar D^4\bar{{\cal W}}^2)^2 +
3(D^4{\cal W}^2)(\bar D^4\bar{{\cal W}}^2)] \nn
&& -\,\frac23 [\bar{{\cal W}}^3D^4{\cal W}^2\Box{\cal W}^3 + {\cal W}^3(\bar D^4\bar{\cal W}^2)\Box\bar{\cal W}^3]  + \frac1{36}{\cal W}^4\Box^2\bar{{\cal W}}^4\Big\},
\lb{8ord}
\eea
\bea
{I}_{BI}^{(10)} &=&\frac18\int d^{12}Z\Big\{\frac1{4}{\cal W}^2\bar{\cal W}^2
\Big[(D^4{\cal W}^2)^3 + (\bar{D}^4\bar{\cal W}^2)^3
+4(D^4{\cal W}^2)^2 \bar{D}^4\bar{\cal W}^2 \nn
&&
+\,4 (D^4{\cal W}^2)(\bar{D}^4\bar{\cal W}^2)^2 + 2(\bar{D}^4\bar{\cal W}^2) D^4({\cal W}^2
\bar{D}^4\bar{\cal W}^2)
+2(D^4{\cal W}^2)\bar{D}^4(\bar{\cal W}^2
D^4{\cal W}^2)\Big]\nn
&&-\frac1{3}{\cal W}^3\bar{\cal W}^2 (\bar{D}^4\bar{\cal W}^2)\Box\bar{D}^4
\bar{\cal W}^3
-\frac1{3}{\cal W}^2\bar{\cal W}^3 (D^4{\cal W}^2)\Box D^4{\cal W}^3
\nn
&&
-\frac2{9}{\cal W}^3\bar{\cal W}^2 (D^4{\cal W}^2) \Box\bar{D}^4\bar{\cal W}^3
-\frac2{9}{\cal W}^2\bar{\cal W}^3 (\bar{D}^4\bar{\cal W}^2)\Box D^4{\cal W}^3
-\frac4{9} {\cal W}^3 (\bar{D}^4 \bar{\cal W}^2)\Box(\bar{\cal W}^3 D^4{\cal W}^2)
\nn
&&
+\,\frac1{36}{\cal W}^4(\Box\bar{\cal W}^3) \Box\bar{D}^4\bar{\cal W}^3
+\frac1{36}\bar{\cal W}^4(\Box {\cal W}^3) \Box D^4{\cal W}^3\nn
&&+\,\frac1{36}{\cal W}^4 (\bar{D}^4\bar{\cal W}^2)\Box^2\bar{\cal W}^4
+\frac1{36}\bar{\cal W}^4 (D^4{\cal W}^2)\Box^2{\cal W}^4
-\frac1{1800}{\cal W}^5\Box^3\bar{\cal W}^5\Big\}.\lb{10ord}
 \eea
The contributions from the three terms in  \p{splt} are easily recognized here. The terms like $B\Box^k\bar B$ in \p{6ord}-\p{10ord}
are hermitian up to a total derivative.

The straightforward (though rather cumbersome) calculations show that the sum of the free action $S_2$ and the interactions \p{4ord} - \p{10ord}
is invariant, to the given order, under the nonlinear $c$ and $\bar c$ central charge transformations \p{B1}. Moreover, all the terms
in \p{4ord} - \p{10ord} can be uniquely fixed, step by step, from the requirement of invariance under these transformations (actually, under the $c$
transformations, because the $\bar c$ invariance follows automatically as a consequence of the reality of the action).

\subsection{An alternative calculation of ${\cal A}_n$}
So far,  we reminded the basics of the formalism worked out in \cite{BIK} and, as a new result, gave the explicit form of the ${\cal N}=4/{\cal N}=2$ BI
action up to the 10th order
in ${\cal W}, \bar{{\cal W}}$. Now we would like to show that there exists an alternative method of expressing the chiral
superfields ${\cal A}_n$ in terms of
the original ${\cal N}=2$ superfield strengths and their derivatives. Its basic advantage is that it directly yields the correct expressions for
${\cal A}_n, n\geq 0\,,$ and not for $\Box^n{\cal A}_n$, as in the approach based on the constraints \p{B7} - \p{B9} and their higher $n$ generalizations.

Our starting point will be the linear realization of the central charge with the parameter $c$ on the full set of chiral functions ${\cal W},
{\cal A}_n, \,n\geq 1,$ and their conjugates, in accordance with the transformation laws  \p{B1} - \p{B4}. Denoting this central charge generator as $Z$, we write
\bea
&& (\mbox{a})\;Z\,{\cal W}=1-\frac12\bar{D}^4\bar{\cal A}_0,\quad
(\mbox{b})\;Z\,\bar{{\cal W}}=\frac12\Box\bar{\cal A}_0,\lb{B61} \\
&& (\mbox{a})\;Z\,{\cal A}_0=2{\cal W},\qquad \qquad \;(\mbox{b})\,Z\,\bar{{\cal A}}_0 = \frac12\Box \bar{{\cal A}}_1\,, \lb{B62} \\
&& (\mbox{a})\;Z\,{\cal A}_n=2{\cal A}_{n-1},\quad \qquad \,
(\mbox{b})\;Z\, \bar{\cal A}_n =\frac12\Box\bar{\cal A}_{n+1}.\lb{B6}
\eea
The action of the conjugated central charge generator $\bar Z$ corresponding to the transformations with the parameter $\bar c$ can be obtained
by complex conjugation.

Next, we assume that all ${\cal A}_n$ can be covariantly expressed in terms of  ${\cal W}, \bar{\cal W}$, have the perturbative expansions
as in \p{B10}, and that ${\cal A}_0, \bar{\cal A}_0$ start with  ${\cal W}^2$ and $\bar{\cal W}^2\,$
\bea
{\cal A}_0^{(2)} = {\cal W}^2\,, \quad  \bar{\cal A}_0^{(2)} = \bar{\cal W}^2\,.
\eea
Surprisingly, this minimal set of assumptions is sufficient for restoring, by recursions,  the whole set of the perturbative terms in ${\cal A}_n$
by the group relations \p{B61} - \p{B6} adapted to the nonlinear realizations ${\cal A}_n = {\cal A}_n({\cal W}, \bar{\cal W})$.

To this end, we consider the perturbative expansion of the central charge generators $Z, \bar Z$ in the nonlinear realization considered
\bea
&&Z=\partial +\sum\limits_{n=1}^\infty Z^{(2n-1)}\,,\quad \bar Z=\bar\partial
 +\sum\limits_{n=1}^\infty \bar{Z}^{(2n-1)}\,, \lb{nonlZ} \\
&& Z^{(2n-1)}{\cal W}=-\frac12\bar{D}^4\bar{\cal A}_0^{(2n)}({\cal W}),\;\;
Z^{(2n-1)}\bar{\cal W}=\frac12\Box\bar{\cal A}_0^{(2n)}({\cal W}), \quad \mbox{and \; c.c.}\,,\lb{ZWn} \\
&& \partial{\cal W}=1,\;\; \partial\bar{\cal W}=0\,, \qquad
\bar\partial{\cal W}= 0,\;\; \bar\partial\bar{\cal W}= 1\,.\lb{hatdelta}
\eea

The nonlinear parts of the $Z$ generators can be found from the evident requirement that all transformations in \p{B61} - \p{B6} are
now induced by the transformations of ${\cal W}$ and $\bar{\cal W}$ defined in \p{ZWn} and \p{hatdelta}.
In particular, for ${\cal A}_0$ we have
\bea
Z^{(2n-1)}{\cal A}_0^{(2m)}=-\frac12[\bar{D}^4\bar{\cal A}_0^{(2n)}]\,
\partial {\cal A}_0^{(2m)}+\frac12[\Box\bar{\cal A}_0^{(2n)}]\,
\bar\partial {\cal A}_0^{(2m)}. \lb{ZnonlA0}
\eea
Eqs. \p{ZnonlA0} are understood in such a way that the variations
$\partial{\cal W}$ and
$\bar\partial\bar{\cal W}$ appearing inside ${\cal A}_0^{(2m)}$
(and, generically, standing under
the differential operators like $\Box^m, D^4, \bar D^4\,$)
are just replaced by the nonlinear coefficients within the square brackets defined in \p{ZnonlA0}. This will be illustrated on a few
examples presented below and in the Appendix B.

The perturbative expansion of the first equation in \p{B62}, that is
$$
Z{\cal A}_0({\cal W})=2{\cal W},
$$
reads
\bea
\partial {\cal A}_0^{(2)} = 2 {\cal W}\,,\quad
\partial{\cal A}_0^{(2n)}+Z^{(2n-3)}{\cal W}^{2}+\sum\limits_{m=2}^{n-1}Z^{(2n-2m-1)}{\cal A}_0^{(2m)}
=0\,, \quad n\geq 2\,. \lb{ChiralEqs}
\eea
The first chiral equation is identically satisfied, while the second one
yields an infinite set of the recursion relations for determining ${\cal A}_0$.
Several first relations are as follows
\bea
&&\partial{\cal A}_0^{(4)} + Z^{(1)}{\cal A}_0^{(2)}=0\,,\quad
\partial{\cal A}_0^{(6)} + Z^{(3)}{\cal A}_0^{(2)}+Z^{(1)}{\cal A}_0^{(4)}=0,
\lb{rec1} \\
&&\partial{\cal A}_0^{(8)} + Z^{(5)}{\cal A}_0^{(2)}+Z^{(3)}{\cal A}_0^{(4)}
+Z^{(1)}{\cal A}_0^{(6)} =0\,.\lb{rec2}
\eea
Explicitly, the first equation in \p{rec1} is
\bea
\partial{\cal A}_0^{(4)} = {\cal W} \bar D^4 \bar{\cal W}^2.
\eea
In view of the definition \p{hatdelta}, ${\cal A}_0^{(4)}$ is just primitive
of the r.h.s. with respect to the argument ${\cal W}$:
\bea
{\cal A}_0^{(4)}\equiv \int_{{\cal W}}\partial{\cal A}_0^{(4)}
= \int_{{\cal W}} {\cal W} \bar D^4 \bar{\cal W}^2 = \frac12 {\cal W}^2
\bar D^4 \bar{\cal W}^2\,,\nonumber
\eea
that coincides with the relevant expression in \p{A046}. Having at hand
${\cal A}_0^{(4)}$, it is easy to calculate the terms
$$Z^{(3)}{\cal A}_0^{(2)}=2{\cal W}Z^{(3)}{\cal W},\quad
 Z^{(1)}{\cal A}_0^{(4)}={\cal W}(\bar{D}^4\bar{\cal W}^2) Z^{(1)}{\cal W}
 +{\cal W}\bar{D}^4(\bar{\cal W}Z^{(1)}\bar{\cal W})$$
 and to find the explicit
expression for $\partial{\cal A}_0^{(6)}$
\bea
\partial{\cal A}_0^{(6)}= \frac12\bar{D}^4\Big[{\cal W}\bar{\cal W}^2
(D^4{\cal W}^2 + \bar D^4\bar{\cal W}^2)
- {\cal W}^2 \bar{\cal W}\Box\bar{\cal W}^2\Big]. \lb{A6prim}
\eea
It is also rather easy to find the primitive of this expression
(using, at one of the intermediate steps, the identity \p{A10}).
The answer coincides with the corresponding expression in \p{A046}.

At this step we encounter an ambiguity. The primitive of the first term in \p{A6prim} ($\sim {\cal W} D^4 {\cal W}^2$) is defined up to the
``integration constant''
\bea
C = {\cal W}^2 D^4{\cal W}^2 -\frac23{\cal W}^3D^4{\cal W} -\frac23{\cal W}D^4{\cal W}^3 + \frac16 D^4{\cal W}^4\,, \quad \partial C = 0\,.\lb{IntC}
\eea
However, this uncertainty is fully fixed by recalling that ${\cal A}_0^{(6)} = 2\bar D^4 L^{(6)}$ (eq. \p{defIBI}), where $L^{(6)}$
is real up to a total derivative (eq. \p{ReaL}).
The contribution of $C$ \p{IntC} is not compatible with this reality property and so should be discarded, leaving us with the expression \p{A046}
manifestly satisfying this reality criterion. Another way to see that $C$ does not contribute is to check the validity of the equation with $\Bar Z$
\bea
\bar\partial{\cal A}_0^{(6)} + \bar Z^{(3)}{\cal A}_0^{(2)}+ \bar Z^{(1)}{\cal A}_0^{(4)}= \frac12 \Box {\cal A}_1^{(5)}\,,\lb{BarEq}
\eea
where ${\cal A}_1^{(5)}$ is defined in \p{A013}. Once again, this equation requires that the coefficient before the possible contribution of $C$ be vanishing.

The more direct way to avoid the ambiguities of this type is to rewrite the second equation in \p{ChiralEqs} as the equation for $L^{(2n)}$
\bea
&&
\partial L^{(2n)}-\frac12{\cal W}\bar{\cal A}_0^{(2n-2)}
+\sum\limits_{m=2}^{n-1}Z^{(2n-2m-1)}L^{(2m)}
=0 \lb{ZnonlL0}\,,
\eea
which is obtained by expressing ${\cal A}_0^{(2n)} = 2\bar D^4L^{(2n)}, \; Z^{(2n-3)} = -\frac12 \bar D^{4} \bar {\cal A}_0^{(2n -2)}\partial$ in \p{ChiralEqs}
and taking off the operator $\bar D^4$. Because of the reality of $L^{(2n)}$, the $\bar Z$ equation does not yield any new information. So the single eq. \p{ZnonlL0}
uniquely specifies $L^{(2n)}$ and, hence, the whole $L_{BI}$. It is remarkable that for this recursion computation of $L^{(2n)}$ one needs
to know only $L^{(2m)}$ (and ${\cal A}_0^{(2m)}= 2\bar D^4 L^{(2m)}$)
with $m \leq n$ and does not need to know ${\cal A}_n, n \geq 1$. It is easy to reproduce the correct $L^{(6)}$ and $L^{(8)}$ in this way,
so that the expressions for
${\cal A}_0^{(6)} = 2\bar D^4L^{(6)}$ and ${\cal A}_0^{(8)} = 2\bar D^4L^{(8)}$ coincide with those given in \p{A046} and \p{A08}. The
expression for ${\cal A}_0^{(10)}({\cal W}, \bar{\cal W})$, eqs. \p{divid} - \p{Y10}, can also be re-derived by making use of \p{ZnonlL0}.
Calculating the next-order terms in ${\cal A}_0$ is also feasible, though such a computation gets more and more involved with each new recursion.

Some other examples of applying the alternative approach with the differential $\partial$ and $\bar\partial$ equations are given in Appendix B.
In particular, it can be used for the recursion
calculation of the higher $n$ chiral superfields, based on the relation
(\ref{B6}a), with the same realization
of $Z^{(2n-1)}$ as in \p{ZnonlA0} (where one should replace
${\cal A}_0^{(2m)} \rightarrow  {\cal A}_p^{(p +2m)},\, p\geq 1\,$). In this way, the generic formulas can be obtained for
the first two terms in the perturbative expansion of ${\cal A}_n$:
\bea
&&{\cal A}_n^{(n+2)}({\cal W})=\frac{2^{n+1}}{(n+2)!}{\cal W}^{n+2},\nn
&&{\cal A}_n^{(n+4)}({\cal W})=\frac{n+1}2{\cal A}_n^{(n+2)} \bar{D}^4\bar{\cal W}^2
=\frac{2^{n}(n+1)}{(n+2)!}{\cal W}^{n+2} \bar{D}^4\bar{\cal W}^2\,.\lb{Areg}
\eea
These expressions are deduced by successively integrating the chains of equations
\bea
\partial{\cal A}_n^{(n+2)} = 2 {\cal A}_{n-1}^{(n+1)}\,, \quad
\partial{\cal A}_n^{(n+4)} -\frac12 \bar{D}^4\bar {\cal W}^2\,
\partial{\cal A}_n^{(n+2)}
=  2 {\cal A}_{n-1}^{(n+3)}\,,
\eea
which follow from (\ref{B62}a) and (\ref{B6}a). For the higher-order ${\cal A}_n^{(n + 2p)}, p\geq 3\,,$ integration of the
corresponding $\partial$ equations can produce the ``integration constants'' like \p{IntC} and, for selecting unambiguous solutions,
one would be forced to resort to the $\bar\partial$ equations like \p{BarEq}\footnote{For the functions \p{Areg} the $\bar\partial$
equations are satisfied identically.}. This is just the case for the examples of Appendix B.

It is curious that the considerations based solely upon the realization
of the central charges $Z$ and $\bar Z$ on the superfields
${\cal W}, \bar{\cal W}$ and ${\cal A}_n$ (eqs. \p{B61}, (\ref{B62}), (\ref{B6}) and their conjugates) yield the correct expressions
for ${\cal A}_n$ without any use of the original set of constraints.
Moreover, for restoring  ${\cal A}_n$ in a given order one needs to know only
the expressions for the lower orders in ${\cal A}_n$,
as well as in all ${\cal A}_p$ with $p<n$.
As was already mentioned, the basic superfield
${\cal A}_0$ can be restored order by order, using solely the $\partial$ equations, without
any need to apply to ${\cal A}_n, n\geq 1\,$.
Nevertheless, to have the full set of
chiral superfields seems to be necessary for checking
the consistency with the nonlinearly realized ${\cal N}=4/{\cal N}=2$
supersymmetry. Indeed,
the transformations  \p{B1} - \p{B4} also imply the validity of the Grassmann-odd equations corresponding
to the mutually conjugated $\xi^k_\alpha$ and $\bar\xi_k^{\dot\alpha}$  transformations. However, these additional relations look
as the consistency conditions  for  the basic ones. It is easy, e.g., to check their validity
for a few first  perturbative terms in $\bar{\cal A}_0$ and $\bar{\cal A}_n, n\geq 1$. Note that the similar conclusions about
the specific interplay between the restrictions following from the central charge symmetry and broken supersymmetry were made in \cite{BIK} in
the perturbative approach exploiting the constraints \p{B7} - \p{B9} and their higher $n$ counterparts. Though the precise relationship
between the two approaches for the time being is not obvious to us, they both result in the same final answers and so are expected
to be equivalent.

To shed more light on the interplay between the central charge and ${\cal N}=4$ supersymmetry invariances, let us explicitly
write some restrictions following from the left nonlinear supersymmetry in \p{B1} - \p{B4}  (corresponding to the parameters $\xi^\alpha_k$).
Denoting the relevant odd generator $S^k_\alpha$, we find
\bea
S^k_\alpha {\cal W} = -2i\theta^k_\alpha Z{\cal W}\,, \quad S^k_\alpha \bar{\cal W} = -2i\theta^k_\alpha Z\bar{\cal W} +
\frac12 \bar D^{k\dot\alpha}\partial_{\alpha\dot\alpha}\bar{\cal A}_0\,, \nonumber
\eea
where the additional term in the transformation of $\bar{\cal W}$ guarantees both sides to be anti-chiral. We observe that the realization of
the left supersymmetry on ${\cal W}, \bar{\cal W}$ is almost completely specified by the realization of the central charge $Z$. The action
of this supersymmetry on the chiral Lagrangian density ${\cal A}_0$ is in fact fully determined by $Z$:
\bea
S^k_\alpha {\cal A}_0 = - 2i \theta^k_\alpha Z {\cal A}_0 = -4i \theta^k_\alpha {\cal W}\,.
\eea
Thus the $S^k_\alpha$ invariance of the chiral integral $\int d^8{\cal Z}\,{\cal A}_0$ follows from its $Z$ invariance. The conjugated
antichiral integral $\int d^8\bar{\cal Z}\,\bar{\cal A}_0$ is manifestly invariant under the $\bar Z$ and $\bar S_{k\dot\alpha}$
transformations. The reality of these chiral integrals (since their Lagrangians are coincident and real up to a total derivative) guarantees the
entire ${\cal N}=4$ supersymmetry of the ${\cal N}=4/{\cal N}=2$ BI action.

\setcounter{equation}0
\section{${\cal N}=2$ BI action with auxiliary chiral superfields}
Now we are prepared to turn to our basic aim, that is constructing a new formulation of the ${\cal N}=2$ BI actions in terms of
the auxiliary superfields.

\subsection{${\cal N}=2$ self-duality and auxiliary superfields}
The formalism of auxiliary (super)fields
gives the general description of the self-dual theories of nonlinear electrodynamics and its
superextensions. Auxiliary bispinor fields in the nonlinear
electrodynamics were considered in \cite{IZ,IZ2,IZ3}, the
auxiliary chiral spinor ${\cal N}=1$ superfields were introduced in \cite{ILZ,Ku}. The similar auxiliary
chiral scalar superfields were also used to construct the actions of the ${\cal N}=2$ self-dual theories
\cite{Ku,BCFKR,CarKa}.

Introducing the auxiliary chiral scalar superfield ${\cal U}\,$, we can consider the
following extended bilinear action:
\bea
&&{\cal S}_{b}({\cal W},{\cal U})=\int d^8{\cal Z}\,{\cal L}_{b}({\cal W},{\cal U})
+\mbox{c.c.},\lb{L2WU1}\\
&&{\cal L}_{b}({\cal W},{\cal U})=-\frac12{\cal U}^2+{\cal U}{\cal W}
-\frac14{\cal W}^2=\frac14{\cal W}^2-\frac12({\cal U}-{\cal W})^2\,.
\lb{S2WU}
\eea

The ${\cal W}$ equation of motion reads\footnote{We vary with respect
to ${\cal W}$ and ${\cal U}$
as independent chiral superfields. The Bianchi identity \p{A9} is
imposed afterwards.}
\bea
&&D^{kl}{\cal M}({\cal W}, {\cal U})-\bar
D^{kl} \bar{\cal M}(\bar {\cal W}, \bar{\cal U})=0,\lb{Weq}\\
&&
{\cal M}({\cal W},{\cal U})=-2i\frac{\delta{\cal S}_b}{\delta {\cal W}}
=-2i\frac{\partial{\cal L}_b}{\partial {\cal W}}
=i({\cal W}-2{\cal U})\,. \lb{UFM}
\eea

The equation of motion for the auxiliary superfield is just
\be
{\cal U}={\cal W}\,, \lb{freeUW}
\ee
and substituting this back into \p{S2WU} and \p{UFM} yields the standard
${\cal N}=2$ bilinear action \p{free1} and the free equation of motion.

In  the $({\cal W},{\cal U})$ representation, the general action of the nonlinear ${\cal N}=2$ electrodynamics ${\cal S}({\cal W},{\cal U})$
is the sum
\be
{\cal S}({\cal W},{\cal U})=
{\cal S}_{b}({\cal W},{\cal U})+{\cal I}({\cal U})\,, \quad
{\cal I}({\cal U}) = \int d^{8}{\cal Z}\, {\cal L}({\cal U})
 + \mbox{c.c.}\,,
 \lb{genintUW}
\ee
where the interaction ${\cal L}({\cal U})$ is a function of
${\cal U}, \bar{\cal U}$ and the composite superfields
which one can construct from ${\cal U}, \bar{\cal U}$ and their $x$ and
$\theta$ derivatives.

Once again, eliminating ${\cal U}, \bar{\cal U}$
by their equations of motion (in the generic case, recursively),
\bea
{\cal U} = {\cal W} + \frac{\delta{\cal I}}{\delta {\cal U}}\,, \qquad
\frac{\delta{\cal I}}{\delta {\cal U}} := {\cal J}({\cal U},
\bar{\cal U}) = \bar D^4 J ({\cal U}, \bar{\cal U})\,, \lb{eqsInt}
\eea
we arrive at the standard nonlinear ${\cal N}=2$ electrodynamics action
\bea
&&S({\cal W}) = S_2({\cal W}) + I({\cal W}) = {\cal S}({\cal W},{\cal U}({\cal W})), \lb{stanD} \\
&& I({\cal W}) = {\cal I}({\cal U})
-\frac12\left[\int d^8{\cal Z}\,\left(\frac{\delta {\cal I}}{\delta {\cal U}}\right)^2 + \mbox{c.c.}  \right]. \lb{IntUW}
\eea
Note the useful equation which directly relates ${\cal U}({\cal W})$ to the ${\cal W}, \bar{\cal W}$ action \p{stanD}:
\bea
{\cal U}({\cal W}) = \frac12 {\cal W} + \frac{\delta{S({\cal W})}}{\delta {\cal W}} = {\cal W} + \frac{\delta{I({\cal W})}}{\delta {\cal W}}\,.\lb{UviaS}
\eea

The ${\cal N}=2$ self-duality condition and the corresponding $U(1)$ duality
transformations in the standard ${\cal W}, \bar{\cal W}$ representation
are given by eqs. \p{A15} and \p{O2} of the Appendix A. In the
$({\cal U}, {\cal W})$ representation, the self-duality amounts to the
off-shell invariance
of the auxiliary interaction ${\cal I}({\cal U})$ under the following $U(1)$ transformations
\be
\delta_\omega
{\cal U}=-i\omega {\cal U}~,\quad\delta_\omega \bar{\cal U}=i\omega \bar{\cal U}\,, \lb{U1U}
\ee
$\omega$ being a real constant parameter. The $U(1)$ transformations of ${\cal W}$ and $\bar{\cal W}$ are
\be
\delta_\omega {\cal W}=\omega\,{\cal M}({\cal W},{\cal U})=
i\omega\,({\cal W}-2{\cal U})\,, \quad
\delta_\omega \bar{\cal W}=\omega\, \bar{\cal M}({\cal W},{\cal U})=
-i\omega\,(\bar{\cal W}-2\bar{\cal U})\,. \lb{U1W}
\ee
Together with \p{U1U}, they ensure the $U(1)$ duality covariance of the
relevant equations of motion for ${\cal W}, \bar{\cal W}\,$
combined with the Bianchi identities \p{A9}. In the interaction case the
dual superfield strength ${\cal M} ({\cal W}, {\cal U})$  and
the dynamical equations of motion are given by the same eqs. \p{UFM} and \p{Weq} as in
the free case.  The specificity of the given nonlinear
system is encoded in the auxiliary equation \p{eqsInt}, i.e. in the
structure of the superfield $J({\cal U}, \bar{\cal U})$.

The $U(1)$ invariance of ${\cal I}({\cal U})$ is equivalent to the integral self-duality condition \p{sdint1}, which in the $({\cal W},{\cal U})$
formulation is reduced to
\bea
\int d^8{\cal Z}\, ({\cal W} {\cal U}-{\cal U}^2) =
\int d^8\bar{\cal Z}\, (\bar{\cal W} \bar{\cal U}-\bar{\cal U}^2)\,.\lb{intinv}
\eea
Using in \p{intinv} the auxiliary equation \p{eqsInt}, we reduce this condition to
\bea
\int d^8{\cal Z}\,{\cal U}\frac{\delta{\cal I}}{\delta {\cal U}} =
\int d^8\bar{\cal Z}\,\bar{\cal U}\frac{\delta{\cal I}}{\delta \bar{\cal U}}\,,
\lb{intinv1}
\eea
which is just the condition of the invariance of the functional
${\cal I}({\cal U})$ under the $U(1)$ transformations \p{U1U}. The general
self-dual
${\cal N}=2$ action admits the representation which is ${\cal N}=2$ analog of
the well-known Gaillard-Zumino representation \cite{GZ} for the bosonic
self-dual actions:
\bea
&& {\cal S}({\cal W},{\cal U})= \frac{i}{4}\int d^8{\cal Z}\,{\cal W}{\cal M} -
\frac{i}{4}\int d^8{\cal Z}\,\bar{\cal W}\bar{\cal M} + {\cal I}'({\cal U})\,,
\lb{GZN2}\\
&&{\cal I}'({\cal U}) = {\cal I}({\cal U})
-\frac12\Big[\int d^8{\cal Z}\,{\cal U}({\cal W} -{\cal U}) + \mbox{c.c.} \Big].
\lb{Iprime}
\eea
The additional term in \p{Iprime} is invariant under \p{U1U} and \p{U1W} on
its own.

Any self-dual system of ${\cal N}=2$ electrodynamics can be reformulated as a
system with the off-shell action \p{genintUW} in which the interaction part
${\cal I}({\cal U})$
is invariant under the $U(1)$ duality transformations \p{U1U}. Conversely, if
some system of ${\cal N}=2$ electrodynamics admits such a reformulation, it is
self-dual.
The conjecture that the ${\cal N}=4/{\cal N}=2$ BI system is self-dual was put
forward for the first time in \cite{KT}. In \cite{BIK} it was proved up
to the 8th order. One of the ways to prove this for the whole ${\cal N}=2$ BI action
is to put the latter into the $({\cal U}, {\cal W)}$ form and to
demonstrate that the corresponding ${\cal I}_{BI}({\cal U})$ is $U(1)$ invariant. Below we make a few steps towards this goal.

\subsection{The $({\cal U}, {\cal W})$ form of the ${\cal N}=2$ BI action:
general structure}

The splitting \p{splt} suggests the following natural conjecture for the
$({\cal W},{\cal U})$ form of the total ${\cal N}=4/{\cal N}=2$ BI action
\bea
&&{\cal S}_{BI}({\cal W},{\cal U})={\cal S}_b({\cal W},{\cal U})
+{\cal I}_{BI}({\cal U})=\int d^8{\cal Z}\,{\cal L}_{BI}({\cal W},{\cal U})
+\mbox{c.c.\,},\\
&&
{\cal I}_{BI}({\cal U})={\cal I}_{\cal X}({\cal U})+{\cal I}_{\cal R}({\cal U}) + {\cal I}_{\cal Y}({\cal U})\,.
\lb{SUBI}
\eea
Here, ${\cal L}_{BI}({\cal W},{\cal U})$ is the full chiral Lagrangian density, and
the interaction functional ${\cal I}_{BI}({\cal U})$ consists of the three different terms.
Below we explain the motivations for including these terms.

The interaction ${\cal I}_{\cal X}({\cal U}) = \int d^8{\cal Z}\,{\cal L}_{\cal X}({\cal U}) +\mbox{c.c.\,}$ in the absence of other terms
should generate just the action associated with the chiral
superfield ${\cal X}$ defined by eq. \p{ketov}. It was proven in \cite{KT} that the corresponding nonlinear ${\cal N}=2$ action is self-dual and defines
an extension of the bosonic BI action. In the next subsection we will give a simpler proof of its self-duality by constructing
the $({\cal U}, {\cal W})$ representation for it.

The next term, ${\cal I}_{\cal R}({\cal U}) = \int d^8{\cal Z}\,{\cal L}_{\cal R}({\cal U}) +\mbox{c.c.\,}$, corresponds to the choice
\bea
{\cal L}_{\cal R} = \frac12 \bar{D}^4\sum_{n=3}^\infty (-1)^{n} \frac{1}{(n!)^2} {\cal U}^n\Box^{n-2}\bar{{\cal U}}^n\,.\lb{RU}
\eea
It is the only structure capable to produce the ${\cal R}$ contribution \p{Rgen} in the ${\cal W}$ representation, taking into account that,
in the lowest order,
${\cal U} = {\cal W} +{\cal O}({\cal W}, \bar{\cal W})$. Note that the first terms in this sum were studied in \cite{BCFKR}
using the approach based on the ``deformed twisted self-duality constraint'', which is equivalent to our approach with auxiliary (super)fields.

Finally, the third term ${\cal I}_{\cal Y}({\cal U}) = \int d^8{\cal Z}\,{\cal L}_{\cal Y}({\cal U}) +\mbox{c.c.\,}$ should be responsible
for possible corrections to the contributions of the first two terms to the total BI action $I_{BI}({\cal W},\bar{\cal W})$. For the time being,
we are not aware of any regular method of constructing such action. We will see that it gives a non-zero contribution starting from the 10th order.

The auxiliary equation of motion \p{eqsInt} for the case under consideration can be written as
\bea
{\cal U} = {\cal W} + \frac{\delta{\cal I}_{BI}}{\delta {\cal U}}\,, \qquad
\frac{\delta{\cal I}_{BI}}{\delta {\cal U}} = {\cal J}_{BI} = \bar D^4 J_{BI} = \bar D^4 (J_{\cal X} + J_{\cal R} + J_{\cal Y})\,. \lb{eqsIntBI}
\eea
Since the whole ${\cal L}_{\cal R}$ is defined by eq. \p{RU}, we can write the full expression for $J_{\cal R}$:
\bea
{J}_{\cal R} = \sum_{n=3}^\infty (-1)^{n} \frac{1}{(n!)(n-1)!} {\cal U}^{n-1}\Box^{n-2}\bar{{\cal U}}^n\,.\lb{RJ}
\eea
We still have no closed expressions for other two terms in $J_{BI}$. For the time being, we know ${\cal I}_{\cal X}({\cal U})$ up to the 16th order
and ${\cal I}_{\cal Y}({\cal U})$ up to the 10th order (see next sections).

\setcounter{equation}{0}
\section{\lb{3}Simple self-dual ${\cal N}=2$ model}
Here we consider in some detail the self-dual model associated with the superfield chiral density ${\cal X}$ in \p{splt} as a subsector of
the full hypothetical ${\cal N}=2$ BI model.

\subsection{New auxiliary superfield formulation}
The superfield ${\cal N}=2$ action ${S}_{\cal X}({\cal W})  = {S}_2({\cal W}) + {I}_{\cal X}({\cal W})$
is the minimal self-dual ${\cal N}=2$ superextension  of the bosonic BI action. It was constructed  in \cite{Ke}
\bea
&&{S}_{\cal X}=\frac14\int d^8{\cal Z}{\cal X}({\cal W})+\frac14\int d^8
\bar{\cal Z}\bar{\cal X}({\cal W}),\lb{Xact}
\eea
where the chiral auxiliary superfield satisfies the simple constraint
\bea
&&{\cal X}={\cal W}^2+\frac12{\cal X}\bar{D}^4\bar{\cal X}.\lb{Xchir}
\eea
The self-duality of ${S}_{\cal X}$ was demonstrated in \cite{KT}.
The  perturbative solution for
$${\cal X}({\cal W})=\sum\limits_{n=1}^\infty
{\cal X}^{(2n)}= {\cal W}^2 + 2\bar{D}^4\sum\limits_{n=1}^\infty
L_{\cal X}^{(2n)}$$
 and the corresponding superfield densities in the full superspace,
$L_{\cal }^{(2n)}({\cal W})\,$,
 were constructed, up to the 8th order in ${\cal W},\bar{\cal W}$,  in \cite{Ke,KT} and, up to the 14th order, in \cite{Bel}.
 Up to the 10th order, the corresponding interaction $I_{\cal X}({\cal W})$ can be obtained by neglecting  all terms with the operator $\Box$ in the
 sum of the actions defined in  eqs. \p{4ord} - \p{10ord} (the chiral density ${\cal X}$ to the same order is a sum of \p{ket10}
and the lower-order terms singled out from eqs. \p{A046} and \p{A08}).

As the preparatory step for passing to the $({\cal U},{\cal W})$ formulation of this model, we
will present a different auxiliary superfield formalism for it, which enables writing its
action in a closed form. The new representation is an analog of the similar formalism developed in \cite{ILZ} for
self-dual ${\cal N}=1$ gauge models.

First, we introduce the constraint \p{Xchir} into the action with the help of
the Lagrange multiplier $R$:
\bea
&&\tilde{S}_{\cal X}({\cal W}, {\cal X}, M)=\frac14\int d^8{\cal Z}{\cal W}^2+
\frac14\int d^8\bar{\cal Z}\bar{\cal W}^2+\frac14\int d^{12}Z{\cal X}\bar{\cal X}
\nn
&&+\frac14\int d^{12}Z\{\bar{R}[{\cal W}^2+\frac12\bar{D}^4({\cal X}
\bar{\cal X})-{\cal X}]
+R[\bar{\cal W}^2+\frac12D^4({\cal X}\bar{\cal X})-\bar{\cal X}]\}, \lb{Xact2}
\eea
where ${\cal X}$ and $R$ are some complex auxiliary superfields.
Varying this extended action with respect to $\bar{R}\,$,
we obtain the constraint
\be
{\cal X}={\cal W}^2+\frac12\bar{D}^4({\cal X}\bar{\cal X})\,,\lb{X2const}
\ee
which has the same chiral perturbative solution as the equation
for ${\cal X}$ \p{Xchir}. Note that the chirality of ${\cal X}$ arises
as a consequence of eq. \p{X2const},
while in the action \p{Xact2} this superfield is {\it unconstrained}, like
the superfield $R\,$.
Substituting the solution of \p{X2const} into $\tilde{S}_{\cal X}\,$, we come back to
the original action \p{Xact}.

On the other hand, varying \p{Xact2} with respect to
$\bar{\cal X}\,$, we obtain the equation
\bea
{\cal X}-R+\frac12{\cal X}\bar{D}^4\bar{R}+\frac12{\cal X}D^4R \lb{XReq}
=0\,.
\eea
Eliminating ${\cal X}$ with the help of this equation,
\bea
{\cal X}(R)=\frac{R}{1+\frac12 r+\frac12\bar r},\qquad
r=\bar{D}^4\bar{R},\quad\bar r=D^4R\,,\nonumber
\eea
and substituting ${\cal X}({R})$ back into \p{Xact2}, we find
the equivalent $({\cal W}, R)$ representation  of the action $S_{\cal X}$
\bea
S_{\cal X}({\cal W}, R) &=& \frac14\int d^8{\cal Z}{\cal W}^2(1+ r)+
\frac14\int d^8\bar{\cal Z}\bar{\cal W}^2(1+\bar r)
+\frac14\int d^{12}Z\, I(R)\,,\lb{SWR}\\
I(R) &=& -\frac{R\bar{R}}{(1+\frac12 r+\frac12\bar r)}\,
.\lb{IR}
\eea
The auxiliary equation for this action (obtained by varying with respect to $\bar{R}$)
is again equivalent to \p{X2const}
\bea
{\cal W}^2-{\cal X}(R)
+\frac12\bar{D}^4\left[{\cal X}(R)\bar{\cal X}(R)\right]=0\,. \lb{XRconst}
\eea
Using this equation in the action \p{SWR}, one can reduce the latter, modulo
a total derivative,
to \p{Xact}\footnote{The simplest way to accomplish this is to trade
${\cal W}^2, \bar{\cal W}^2$
for ${\cal X}, \bar{\cal X}$ by eq. \p{XRconst}.}.

Thus we derived a new off-shell formulation for the considered system in terms
of the ${\cal N}=2$
superfield strengths ${\cal W}, \bar{\cal W}$ and a complex
{\it unconstrained} auxiliary ${\cal N}=2$ superfield $R$.
In this formulation the action has the closed form \p{SWR}, \p{IR}.
The previously known representation for the action as
an infinite series in ${\cal W}, \bar{\cal W}$ and their derivatives arises after
elimination
of $R$ by its equation of motion.

Using the relations
\bea
&&{\cal M}({\cal W}, R)=-2i\frac{\delta S({\cal W}, R)}{\delta {\cal W}}=
-i{\cal W}(1+ r)
,\nn
&&\delta_\omega{\cal W}=\omega\, {\cal M}({\cal W}, R)\,,
\eea
we find how the duality transformations \p{O2} look in this particular model
\bea
&&\delta_\omega{\cal W} =-i\omega\,{\cal W}(1+ r),\nn
&&\delta_\omega\bar{R}=2i\omega\,(1+\frac12 r)\bar{R},\quad
\delta_\omega r=2i\omega\,(1+\frac12 r)r. \lb{Rduality}
\eea
The transformation of $R$ is uniquely fixed by requiring that
$\delta {\cal M}({\cal W}, R) = -\omega\, {\cal W}$.
The auxiliary interaction  \p{IR} is invariant under these transformations,
while the remaining terms in \p{SWR}
can be rewritten by analogy with the ${\cal N}=2$ GZ representation
 \p{GZN2}
\bea
S_{\cal X}({\cal W},R)=\frac{i}4\int d^8{\cal Z}\,{\cal W}
{\cal M}({\cal W}, R)
-\frac{i}4\int d^8\bar{\cal Z}\,\bar{\cal W}\bar{\cal M}({\cal W}, R)
+\frac14\int d^{12}Z\,I(R).
\eea
The existence of such a representation for the general ${\cal N}=2$  gauge model
superfield action amounts
to the self-duality condition \p{A15}. Thus the model under consideration is self-dual,
in agreement with the conclusion drawn in  \cite{KT}.

\subsection{Passing to the $({\cal U}, {\cal W})$ formulation}
In order to derive the equivalent $({\cal U}, {\cal W})$ representation of the model, we need to rewrite \p{SWR} and \p{IR}
in a slightly different form.
We introduce the new auxiliary superfield variables $\bar{N}$ and $n=\bar{D}^4\bar{N}$ related to ${R}, r$ as
\bea
\bar{R}=\frac{\bar{N}}{1-\frac12n},\quad r=\bar{D}^4\bar{R}
=\frac{n}{1-\frac12n}\,.\lb{RN}
\eea
Their nice property is that the duality transformations \p{Rduality} act on
them linearly,
\bea
&&\delta_\omega\bar{N}=2i\omega\,\bar{N},\quad \delta_\omega n=2i\omega\, n\,.
\lb{lindual}
\eea
In terms of the new variables the action \p{SWR} is rewritten as
\bea
S_{\cal X}({\cal W},N)=\frac14\int d^8{\cal Z}\,{\cal W}^2\frac{1+\frac12n}{1-\frac12n}
+\frac14\int d^8\bar{\cal Z}\,\bar{\cal W}^2\frac{1+\frac12\bar{n}}
{1-\frac12\bar{n}}+\frac14\int d^{12} Z\, {\cal L}_{\cal X}(N), \lb{SWN}
\eea
with
\bea
&&{\cal L}_{\cal X}(N)=-\frac{N\bar{N}}{1-\frac14n\bar{n}}\,, \lb{SWN1}
\eea
by analogy with the corresponding ${\cal N}=1$ case \cite{ILZ}. The equations of motion for the auxiliary
superfields $N, \bar N$ once again yield the chiral constraint
\p{XRconst} which reduces \p{SWN} to \p{Xact}.

The action \p{SWN} is the starting point for finding out the standard
$({\cal W},{\cal U})$ representation for the action of our system:
\bea
&&{\cal S}_{\cal X}({\cal W},{\cal U})={\cal S}_b({\cal W},{\cal U})
+{\cal I}_{\cal X}({\cal U}^2)\,.
\lb{SUX}
\eea

We introduce a chiral ${\cal N}=2$ superfield ${\cal U}$ and write the $({\cal U}, {\cal W}, N)$ image of the action \p{SWN} as
\bea
{\cal S}_{\cal X}({\cal W}, N) \quad \Rightarrow\quad {\cal S}_{\cal X}({\cal W}, {\cal U}, N) =
{\cal S}_b({\cal W},{\cal U}) + {\cal I}_{\cal X}({\cal U}^2, N)\,, \lb{WUN}
\eea
where
\bea
{\cal I}_{\cal X}({\cal U}^2,N) &=& \frac14\int d^{12}Z\left[{\cal U}^2\bar{N}
+\bar{\cal U}^2N
+{\cal L}(N)\right] \nn
&=& \frac14\int d^{12}Z\Big\{{\cal U}^2\bar{N}+\bar{\cal U}^2N
-\frac{N\bar{N}}{1 - \frac14 n\bar n}\Big\}. \lb{UNint}
\eea

Using the ${\cal U}$ equation
\be
{\cal U}=\frac{{\cal W}}{1-\frac12n}\lb{WUneq}
\ee
in the action \p{WUN}, we return to the action \p{SWN}.
On the other hand, varying with respect to $\bar{N}$ (and $N$), we obtain the equation for $N({\cal U}^2)$
\bea
&&N - (1-\frac14n\bar{n}){\cal U}^2 + \frac14 (1-\frac14n\bar{n}) \bar{D}^4\left[\frac{N\bar{N}\bar{n}}{(1-\frac14n\bar{n})^2}\right]
=0\lb{NUeq}
\eea
(and its conjugate). Solving this equation by recursions, we find
\bea
{\cal I}_{\cal X}({\cal U}^2) = {\cal I}_{\cal X}({\cal U}^2, N({\cal U}^2))\,.
\lb{XUint}
\eea
The interaction term \p{UNint} is invariant and the equation \p{NUeq} is covariant with respect to the $U(1)$ duality transformations \p{lindual}, \p{U1U}, so
the ultimate interaction \p{XUint} is also invariant under \p{U1U}. This provides one more proof of the self-duality of the initial model.

For further use, we give a few first recursive solutions of eq. \p{NUeq}:
\bea
N^{(2)} &=& {\cal U}^2,\quad N^{(6)}=-\frac1{4}{\cal U}^2(A+B),\quad
\lb{N26} \\
N^{(10)} &=&
\frac1{16}{\cal U}^2\Big\{B^2+A(2B+\bar{B})+(D^4{\cal U}^2) \bar{D}^4(\bar{\cal U}^2\bar{B})
+(\bar{D}^4\bar{\cal U}^2) D^4({\cal U}^2B) \nn
&&+\,\bar{D}^4D^4\,[{\cal U}^2\bar{\cal U}^2(2\bar{B}+B)]\Big\}, \lb{N10}
\eea
where we denoted
\bea
A :=(D^4{\cal U}^2)(\bar{D}^4\bar{\cal U}^2),\quad B :=
\bar{D}^4D^4(\bar{\cal U}^2{\cal U}^2),
\quad \bar{B}=
D^4\bar{D}^4(\bar{\cal U}^2{\cal U}^2)\,. \nonumber
\eea

Using these solutions, we have constructed few lowest terms of the interaction ${\cal I}_{\cal X}({\cal U}^2)$:
\bea
{\cal I}_{\cal X}({\cal U}^2)
&=& \sum\limits_{n=1}^\infty\int d^{12}Z {\cal L}^{(4n)}_{\cal X}({\cal U},\bar{\cal U}),\nn
{\cal L}^{(4)}_{\cal X} &=& \frac14{\cal U}^2\bar{\cal U}^2,\qquad
 {\cal L}^{(8)}_{\cal X}=-\frac1{16}{\cal U}^2\bar{\cal U}^2\, A,  \lb{L48} \\
{\cal L}^{(12)}_{\cal X} &=& \frac1{64}{\cal U}^2\bar{\cal U}^2\left(B\bar{B}+B^2+\bar{B}^2\right),\lb{C01}\\
{\cal L}^{(16)}_{\cal X} &=& -\frac1{256}{\cal U}^2\bar{\cal U}^2\Big\{D^4[{\cal U}^2(B+2\bar{B})]
\bar{D}^4[\bar{\cal U}^2(\bar{B}+2B)] \nn
&&+\,(B+\bar{B})(B^2+\bar{B}^2)\Big\}.\lb{L16}
\eea
These terms allow one to restore the original action $S_{\cal X}({\cal W})$ up to the 18th order by eliminating ${\cal U}, \bar{\cal U}$
by their equations of motion\footnote{In ref. \cite{Bel}, the action  $S_{\cal X}({\cal W})$ was explicitly given up to the 14th order. In the
${\cal U}$ language, this corresponds to keeping, in \p{L48} - \p{L16}, all terms up to the 12th order.}.

The auxiliary equation for ${\cal S}_{b}({\cal W},{\cal U})+{\cal I}_{\cal X}({\cal U}^2)$
contains the variational derivative
\bea
&&{\cal W}-{\cal U}+2{\cal U}\frac{\delta {\cal I}_{\cal X}}{\delta {\cal U}^2}=
{\cal W}-{\cal U}+\bar{D}^4{J}_{\cal X}\,.\lb{barF}
\eea
Solving the auxiliary  equation for the function ${\cal U}({\cal W})$
 and substituting this solution back into this action,  we
obtain the ``minimal'' ${\cal N}=2$  action $S_{\cal X}({\cal W})$ as an infinite series of the powers of
${\cal W}, \bar{\cal W}$ and their derivatives. Note that this series comprises an enormous  number of terms, such that the
new structures appear with each new recursion \cite{Bel}. In the $({\cal U}, {\cal W})$ formulation, at least  up to the 16th
order, we are left with a limited number of terms which all are expressed through ${\cal U}^2$, $\bar{\cal U}^2$ and
the dimensionless objects $A$ and $B$. This makes it probable that the whole interaction ${\cal I}_{\cal X}({\cal U})$ can be written
as a sum of the well defined  terms related by some general recurrence formula.

\setcounter{equation}0
\section{The $({\cal U}, {\cal W})$ form of the ${\cal N}=4/{\cal N}=2$ BI action
up to 10th order}
Here we will find the auxiliary interaction ${\cal I}_{BI}({\cal U})$ which
reproduces the $({\cal W}, \bar{\cal W})$ form of the
BI action up to the 10th order, i.e. the sum of four terms
\bea
\hat{I}_{BI} = I^{(4)}_{BI} + I^{(6)}_{BI}  + I^{(8)}_{BI} + I^{(10)}_{BI}\,,
\lb{hatI410}
\eea
which were written down in eqs. \p{4ord} - \p{10ord}.

Our starting point will be the general decomposition \p{SUBI} of
${\cal I}_{BI}({\cal U})$ into the three terms.  First, we know that
in order to find the contribution of ${\cal I}_{\cal X}({\cal U})$ to
$I_{BI}({\cal W})$ up to the 10th order, it is enough to keep in
${\cal I}_{\cal X}({\cal U})$ the terms up to the 8th order, i.e. those defined
by eqs. \p{L48}, whence
\bea
\hat{{\cal I}}_{\cal X}({\cal U})  =  \frac14 \int d^{12}Z \,{\cal U}^2\bar{\cal U}^2
\left[1 - \frac14 (D^4{\cal U}^2)(\bar D^4\bar{\cal U}^2)\right].\lb{Xcontr}
\eea

Secondly, we need three terms from ${\cal I}_{R}({\cal U})$ defined by \p{RU}:
\bea
\hat{{\cal I}}_{\cal R}({\cal U}) = \frac18 \int d^{12}Z \,\Big(-\frac{2}{9}{\cal U}^3\Box \bar{\cal U}^3
+ \frac{1}{72} {\cal U}^4\Box^2 \bar{\cal U}^4 -
\frac{1}{1800}{\cal U}^5\Box^3 \bar{\cal U}^5\Big). \lb{Rcontr}
\eea

The sum of the interaction terms \p{Xcontr} and \p{Rcontr} will serve as the input of our construction. They both are invariant under
the duality $U(1)$ group \p{U1U} and so
necessarily yield a self-dual theory after passing to the $({\cal W}, \bar{\cal W})$ representation.
As for possible contributions from ${\cal I}_{Y}({\cal U})$,
they are not known in advance and should be constructed as far as necessary, step by step.

Next we need the explicit expressions for ${\cal U}$ in terms of $({\cal W}, \bar{\cal W})\,$. They can be found by solving
the auxiliary equation \p{eqsIntBI} for $\hat{\cal I} = \hat{{\cal I}}_{\cal X} + \hat{{\cal I}}_{\cal R}$:
\bea
{\cal U} = {\cal W} + \frac{\delta\hat{\cal I}}{\delta{\cal U}} = {\cal W} + \bar{D}^4(\hat{J}_{\cal X}+ \hat{J}_{\cal R})\,.\lb{eqHat}
\eea
The relevant recursion procedure is rather straightforward. Explicitly, eq. \p{eqHat} reads
\bea
{\cal U} &=& {\cal W}+ \frac12\bar{D}^4\Big\{ {\cal U}\bar{\cal U}^2
-\frac1{4}{\cal U}\bar{\cal U}^2[(\bar{D}^4\bar{\cal U}^2) (D^4{\cal U}^2)
+D^4\bar{D}^4({\cal U}^2\bar{\cal U}^2)] \nn
&&-\,\frac{1}{6}{\cal U}^2\Box\bar{\cal U}^3
+\frac1{72}{\cal U}^3\Box^2\bar{\cal U}^4
-\frac1{1440}{\cal U}^4\Box^3\bar{\cal U}^5\Big\} =: {\cal W} + \Delta {\cal U}\,. \lb{Uexpl1}
\eea
The lowest perturbative solutions of this equation are
\bea
{\cal U}^{(1)} = {\cal W}\,, \quad {\cal U}^{(3)} = \frac12\,\bar{D}^4
({\cal W}\bar{\cal W}^2)\,.\lb{U13sol}
\eea

We can calculate the $({\cal W}, \bar{\cal W})$ interaction
$\hat{I}_{BI}({\cal W})$ in two different ways, which yield the same result.
One can find various
orders of these action by directly substituting the perturbative expansion
${\cal U}({\cal W}) = \sum_{n=0}{\cal U}^{(2n+1)}({\cal W})$
into $\hat{\cal S}_{BI}({\cal W},{\cal U}) = {\cal S}_{b}({\cal W}, {\cal U})
+ \hat{\cal I}({\cal U})\,$. Alternatively, one
can make use of the general equation \p{UviaS}, which amounts to
\bea
{\cal U}^{(2n +1)}({\cal W})  = \frac{\delta \hat{I}_{BI}^{(2n + 2)}({\cal W})}
{\delta {\cal W}}\,, \quad n \geq 1\,,\lb{UviaSpart}
\eea
and then calculate $\hat{I}_{BI}^{(2n + 2)}({\cal W})$ by integrating these
equations. To reconstruct $\hat{I}_{BI}$ up to the 10th order,
one needs to know ${\cal U}$ up to the 7th order in the first method, and up to
the 9th order in the second method.

We explicitly quote ${\cal U}^{(5)}$ and  ${\cal U}^{(7)}$ obtained by solving eq. \p{Uexpl1} in the corresponding orders:
\bea
{\cal U}^{(5)} &=& \frac12\bar{D}^4\left[\frac1{2} {\cal W}\bar{\cal W}^2
\bar{D}^4\bar{\cal W}^2
+{\cal W}\bar{\cal W}^2 D^4{\cal W}^2
-\frac1{6}{\cal W}^2\Box\bar{\cal W}^3\right], \lb{B215} \\
{\cal U}^{(7)} &=& \frac14\bar{D}^4\left[\frac1{2} {\cal W}\bar{\cal W}^2(\bar{D}^4\bar{\cal W}^2)^2
+\frac3{2} {\cal W}\bar{\cal W}^2 (\bar{D}^4\bar{\cal W}^2) (D^4{\cal W}^2)
+\frac3{2} {\cal W} \bar{\cal W}^2(D^4{\cal W}^2)^2\right.
\nn
&&\left.
+\,\frac3{2} {\cal W} \bar{\cal W}^2D^4({\cal W}^2 \bar{D}^4\bar{\cal W}^2)
-\frac1{3} {\cal W}\bar{\cal W}^3D^4\Box{\cal W}^3
-\frac1{2}{\cal W}^2(\bar{D}^4\bar{\cal W}^2)\Box\bar{\cal W}^3
\right.\nn
&&\left.
-\,\frac1{2}{\cal W}^2\Box(\bar{\cal W}^3 D^4{\cal W}^2)
+\frac1{36}{\cal W}^3\Box^2\bar{\cal W}^4\right]. \lb{B217}
\eea

After substitution of \p{U13sol}, \p{B215} and \p{B217} into the ``truncated'' $({\cal U}, {\cal W})$ action
\bea
\hat{\cal S}_{BI} = {\cal S}_b + \hat{\cal I}_{\cal X} + \hat{\cal I}_{\cal R}\,,
\eea
and keeping there the ${\cal W}, \bar{\cal W}$ terms up to the 10th order, we obtain the corresponding action
$S({\cal W}, \bar{\cal W})$ to the same 10th order.
The relevant interaction coincides with \p{hatI410} up to the 8th order, but reveals certain deviations from  the correct
10th order term \p{10ord}. These deviations can be fully canceled by adding, to the sum of ${\cal U}$ interactions \p{Xcontr}, \p{Rcontr},
the following 10th order contribution from ${\cal I}_{\cal Y}({\cal U})$:
\bea
\hat{\cal I}_{\cal Y} ({\cal U}) &=& \frac1{72}\int d^{12}Z\,\Big[
{\cal U}^3\bar{\cal U}^2 (D^4{\cal U}^2)\Box\bar{D}^4\bar{\cal U}^3
+{\cal U}^2\bar{\cal U}^3 (\bar{D}^4\bar{\cal U}^2)\Box D^4{\cal U}^3 \nn
&&+\,\frac12{\cal U}^3 (\bar{D}^4\bar{\cal U}^2)\Box(\bar{\cal U}^3 D^4{\cal U}^2) \Big].\lb{Ycontr}
\eea
The last term is hermitian up to a total derivative. The interaction \p{Ycontr} is $U(1)$ invariant,
so it does not break the self-duality of the relevant $({\cal W}, \bar{\cal W})$ action. As a result, we proved that
the 10th order ${\cal N}=2$ BI action $\hat{S}_{BI} = S_2 + \hat{I}_{BI}$ is self-dual.

It is useful to give how the auxiliary equation \p{Uexpl1} is modified upon taking into account the extra interaction \p{Ycontr}:
\bea
\Delta{\cal U} \; &\Rightarrow& \; \Delta{\cal U} + \Delta {\cal U}^{(9)}\,, \lb{Uexpl2} \\
\Delta {\cal U}^{(9)} &=& \frac1{12}\bar D^4 \Big\{\frac1{4}{\cal U}^2 (\bar{D}^4\bar{\cal U}^2)\Box(\bar{\cal U}^3 D^4{\cal U}^2)
+\frac1{6}{\cal U}\bar{\cal U}^3 D^4\Box({\cal U}^3\bar{D}^4\bar{\cal U}^2) \nn
&&+\,\frac12{\cal U}^2\bar{\cal U}^2(D^4{\cal U}^2)\Box\bar{D}^4\bar{\cal U}^3
+\frac13{\cal U}\bar{\cal U}^2D^4({\cal U}^3\Box\bar{D}^4\bar{\cal U}^3)\nn
&&+\,\frac1{3}{\cal U}\bar{\cal U}^3 (\bar{D}^4\bar{\cal U}^2) \Box D^4{\cal U}^3
+\frac1{2}{\cal U}^2 D^4\Box({\cal U}^2\bar{\cal U}^3 \bar{D}^4\bar{\cal U}^2) \Big\}. \lb{AddingU}
\eea
The solutions \p{U13sol}, \p{B215} and \p{B217} are not affected by this modification. Now one can calculate the correct
term ${\cal U}^{(9)}({\cal W})$  and be convinced that eq. \p{UviaSpart} with $n=4$ yields just the expression \p{10ord}
for $\hat I^{(10)}_{BI}({\cal W})$. We will not present details of this straightforward consistency check (which
we have done explicitly to make sure that everything is correct).

Finally, for reader's convenience, we summarize our main results.

We started from the ${\cal N}=2$ BI action describing the spontaneous breaking ${\cal N} =4 \rightarrow {\cal N}=2$ and given, to the 10th order,
by the expression
\bea
\hat{S}_{BI}^{(10)}({\cal W}) = S_2({\cal W}) + I^{(4)}_{BI}({\cal W}) + I^{(6)}_{BI}({\cal W})  + I^{(8)}_{BI}({\cal W})
+ I^{(10)}_{BI}({\cal W})\,, \lb{hatSoncemore}
\eea
where the free part $S_2$  and the interaction terms $I^{(4)}_{BI}$ - $I^{(10)}_{BI}$ are defined by eqs. \p{free1}, \p{4ord} - \p{10ord}.
We showed that this ${\cal N}=2$ BI action  admits,  to the same 10th order in the involved superfields,
the equivalent $({\cal U}, {\cal W})$ formulation as the action
\bea
\hat{\cal S}_{BI}^{(10)}({\cal U}, {\cal W}) = {\cal S}_b({\cal U}, {\cal W}) + {\cal I}^{(4)}_{BI}({\cal U})
+ {\cal I}^{(6)}_{BI}({\cal U})  + {\cal I}^{(8)}_{BI}({\cal U}) + {\cal I}^{(10)}_{BI}({\cal U})\,. \lb{finSUact}
\eea
Here the bilinear part ${\cal S}_b$ is defined by eqs. \p{L2WU1}, \p{S2WU} and the interaction terms
${\cal I}^{(4)}_{BI}$ - ${\cal I}^{(10)}_{BI}$ read
\bea
{\cal I}^{(4)}_{BI} &=& \frac14 \int d^{12}Z\,{\cal U}^2\bar{\cal U}^2\,, \quad {\cal I}^{(6)}_{BI}
= -\frac1{36}\int d^{12}Z\,{\cal U}^3\Box\bar{\cal U}^3,\lb{46ord2} \\
{\cal I}^{(8)}_{BI} &=& -\frac1{16}  \int d^{12}Z\,\Big[{\cal U}^2\bar{\cal U}^2 (D^4{\cal U}^2) (\bar{D}^4\bar{\cal U}^2)
- \frac1{36}{\cal U}^4\Box^2\bar{\cal U}^4\Big], \lb{8ord2} \\
{\cal I}^{(10)}_{BI} &=& \frac1{72}\int d^{12}Z\,\Big\{ {\cal U}^3\bar{\cal U}^2 (D^4{\cal U}^2)\Box\bar{D}^4 \bar{\cal U}^3
+{\cal U}^2\bar{\cal U}^3 (\bar{D}^4\bar{\cal U}^2)\Box D^4{\cal U}^3  \nn
&& +\,\frac1{2}{\cal U}^3 (\bar{D}^4\bar{\cal U}^2)
\Box(\bar{\cal U}^3 D^4{\cal U}^2) -\frac1{200}{\cal U}^5\Box^3\bar{\cal U}^5
\Big\}. \lb{10ord2}
\eea
Eliminating recursively the auxiliary superfield ${\cal U}$ from the action \p{finSUact} by its equation of motion, and keeping
all terms up to the 10th order in ${\cal W}, \bar{\cal W}$, we recover the original truncated BI action \p{hatSoncemore}.
Since the interaction  in the action \p{finSUact} is invariant under the $U(1)$ duality transformations \p{U1U}, the action \p{finSUact} is
a particular case of the duality symmetric ${\cal N}=2$ actions in the $({\cal U}, {\cal W})$ formulation. Hence its ${\cal W}$ representation, i.e.
the truncated BI action \p{hatSoncemore}, also defines a self-dual system.

\setcounter{equation}0
\section{Conclusions and outlook}
In this paper, we studied the possibility that the known ${\cal N}=2$ BI action with the spontaneously broken ${\cal N}=4$
supersymmetry admits the general $({\cal W}, {\cal U})$ representation \p{GenSFN2} with the  $U(1)$ invariant interaction ${\cal I}_{BI}({\cal U})$,
which would mean that this ${\cal N}=2$ BI action indeed defines a self-dual system, as suggested in \cite{KT} and \cite{BIK}. We succeeded
to show this up to the 10th order in the involved superfields. As a by-product, we found the explicit form of the 10th order
of the original BI action  which was known before only up to the 8th order. It is rather straightforward to extend this consideration
to the next, 12th order. All the necessary ingredients for this are already collected in the present paper. In particular, in Appendix B
we present a folded form of the Lagrange density ${L}^{(12)}$ (related to the chiral Lagrangian density ${\cal A}_0^{(12)}$ as
${\cal A}_0^{(12)}= 2\bar D^4 {L}^{(12)}$). However,
we believe that there should exist a method of
proving the self-duality of  the ${\cal N}=4/{\cal N}=2$ BI action to any order, perhaps without inspecting each order step by step.
This hope is based on the fact that the full chiral Lagrangian density ${\cal A}_0$ can be found as the solution of the system of
differential equations  related to the nonlinear realization of the ${\cal N}=4$ central charge on the superfield strengths
${\cal W}, \bar{\cal W}$. This new approach to computing  ${\cal A}_0$ is proposed in section 2.3  of our paper. The ${\cal N}=2$ self-duality, i.e.
$O(2)$ symmetry between the Bianchi identity and nonlinear equation of motion for ${\cal W}, \bar{\cal W}$, could be
a hidden consequence of this basic set of equations. Note that the hypothesis that the central charge (shift) symmetry properly
realized on ${\cal W}, \bar{\cal W}$, being combined with the self-duality requirement, imply invariance under the full spontaneously
broken ${\cal N}=4$ supersymmetry was put forward in ref. \cite{KT}\footnote{See also \cite{BCFKR,CarKa} for a discussion of the interplay between
self-duality and nonlinear supersymmetry.}. Our consideration agrees with this conjecture. It would be also of interest  to reveal possible links with
a recent paper \cite{BCKSV}, where the problem of constructing the full ${\cal N}=2$ BI action with spontaneously broken ${\cal N}=4$ supersymmetry
was treated at the component level and the necessity of deformation of the linear ${\cal N}=2$ supersymmetry was argued.

An interesting open problem is whether the 10th order-truncated $({\cal U}, {\cal W})$ BI action \p{finSUact} can be somehow promoted to all orders in the
auxiliary superfields ${\cal U}, \bar{\cal U}$, thus providing the $({\cal U}, {\cal W})$ form of the complete ${\cal N}=4/{\cal N}=2$ BI
action (still unknown in the closed form). While the structure of the pieces ${\cal I}_{\cal X}$ and ${\cal I}_{\cal R}$
in the general triad decomposition \p{splt} is obvious, it is not true for the part ${\cal I}_{\cal Y}$. We know only that it starts
from the 10th order \p{Ycontr}. It would be tempting to see whether the ${\cal Y}$ terms can be interpreted as perturbative solutions
of some closed superfield equation like, e.g., eq. \p{Xchir}. The closely related problem is to understand how the hidden spontaneously broken
${\cal N}=4$ supersymmetry (including the central charge transformations) is realized in the $({\cal U},{\cal W})$ formulation, i.e.
on the extended superfield set ${\cal W}, \bar{\cal W}, {\cal U}, \bar{\cal U}$. We hope to shed more light on these issues soon.

\section*{Acknowledgements}
We acknowledge a partial support from the RFBR grants Nr.12-02-00517,
Nr.13-02-91330, Nr.13-02-90430, the grant DFG LE 838/12-1 and a grant of the Heisenberg-Landau
program.

\section*{Note added}
After the first version of this paper appeared in Archive, we realized that an important part of ref. \cite{CarKa} directly related
to the subject of our study has escaped our notice\footnote{We thank Stefano Bellucci for bringing  this omission to our attention.}. The authors of \cite{CarKa}
calculated the auxiliary action \p{10ord2} using the equivalent language of ``nonlinear twisted self-duality constraints'' \cite{BN,BCFKR}. However,
only the $\Box^3$ terms in their and our auxiliary actions agree, while the remaining ones do not coincide and there is no way to transform
them into each other. It seems that this discrepancy is related
to the fact that the explicit 10th order ${\cal W}$ action, which was taken as an input in \cite{CarKa}, does not coincide with our \p{10ord},
although it was claimed to follow from ref. \cite{BIK}. We warrant the correctness of \p{10ord} and, hence, of \p{10ord2} because it was checked
in a few independent ways.
So we can conclude that the auxiliary action of \cite{CarKa}, while definitely generating some self-dual ${\cal N}=2$ action, does not reproduce
the 10th order of the genuine ${\cal N}=4/{\cal N}=2$ BI action. For cogency, we added a new Appendix C where some basic steps
leading to \p{10ord} are presented explicitly.

\setcounter{equation}0
\renewcommand{\theequation}{A.\arabic{equation}}
\section*{Appendix A. ${\cal N}=2$ nonlinear electrodynamics}

The ${\cal N}=2, d=4$ superspace in the central basis is parametrized by the
coordinate set $z=(x^m,
\theta^\alpha_k, \bar\theta^{k\dot\alpha})\,$, with the supersymmetry
generators realized as
\bea
Q^k_\alpha=\partial^k_\alpha-i\bar\theta^{k\dot\beta}
(\sigma^m)_{\alpha\dot\beta}\partial_m\,,\quad
\bar{Q}_{k\dot\alpha}=-\bar\partial_{k\dot\alpha}
+i\theta^\beta_k(\sigma^m)_{\beta\dot\alpha}\partial_m\,.
\eea
The ${\cal N}=2$ covariant spinor derivatives are defined by
\bea
D^k_\alpha=\partial^k_\alpha+i\bar\theta^{k\dot\beta}(\sigma^m)_{\alpha\dot\beta}
\partial_m\,,\quad
\bar{D}_{k\dot\alpha}=-\bar\partial_{k\dot\alpha}-i\theta^\beta_k
(\sigma^m)_{\beta\dot\alpha}\partial_m\,.
\eea
All these differential operators  satisfy the relations
\bea
&&\{D_\alpha^k, D^l_\beta\}=\{\bar
D_{k\dot\alpha},\bar D_{l\dot\beta}\}=0~,\quad \{D_\alpha^k,\bar
D_{l\dot\alpha}\}=-2i\delta^k_l
(\sigma^m)_{\alpha\dot\alpha} \partial_m\,,\nn
&&\{Q_\alpha^k, Q^l_\beta\}=\{\bar
Q_{k\dot\alpha},\bar Q_{l\dot\beta}\}=0~,\quad \{Q_\alpha^k,\bar
Q_{l\dot\alpha}\}= 2i\delta^k_l
(\sigma^m)_{\alpha\dot\alpha} \partial_m\,,\nn
&&\{D_\alpha^k, Q^l_\beta\}=\{D_\alpha^k,\bar Q_{l\dot\beta}\} =0\,,\quad
\{\bar D_{k\dot\alpha}, Q^l_\beta\}=\{\bar D_{k\dot\alpha},\bar Q_{l\dot\beta}\} =0\,.
\eea
We use the notation
\bea
&&D^{ik}=D^{\alpha i}D^k_\alpha~,\; D_{\alpha\beta}=D^k_\alpha D_{k\beta}\,,\qquad
\bar{D}_{kl}=\bar D_{k\dot\alpha}\bar D^{\dot\alpha}_l\,,\; \bar
D_{\dot\alpha\dot\beta}=\bar D_{k\dot\alpha} \bar D^k_{\dot\beta}\,,\nn
&&D^4={1\over16}(D^{1\alpha}D^1_\alpha)(D^{2\alpha}D^2_\alpha) = {1\over48}D^{ik}D_{ik} = {1\over48}D^{\alpha\beta}D_{\alpha\beta}\,,\;
D^4\bar{D}^4D^4=\Box^2D^4\,,
\eea
where $\Box = \partial^m\partial_m =\frac12 \partial^{\dot\alpha\alpha}\partial_{\alpha\dot\alpha}\,, \;
\partial_{\alpha\dot\alpha} :=(\sigma^m)_{\alpha\dot\alpha}\partial_m\,$. The chiral and real superspace integration measures are
\bea
d^{12}Z=d^4xD^4\bar{D}^4,\quad
d^8{\cal Z}=d^4xD^4.
\eea

The chiral Abelian ${\cal N}=2$ superfield strengths are defined as
\bea
&&{\cal W}=\bar{D}^4D^{kl}V_{kl}~,\quad \bar{\cal W}=D^4\bar{D}^{kl}V_{kl}\,,
\lb{A7}
\eea
where $V_{kl}$ is the gauge prepotential \cite{Me}
\footnote{The harmonic-superspace description of the
${\cal N}=2$ gauge theory can be found, e.g., in the book \cite{HS}.}.
The corresponding Bianchi identity reads
\be
D^{kl} {\cal W}-\bar D^{kl} \bar {\cal W}=0\,.
\lb{A9}
\ee
The corollary of \p{A9} is the important relations
\bea
&&D^4{\cal W}=-\Box \bar{\cal W}\,, \quad \bar D^4\bar{\cal W}=-\Box {\cal W}\,.\lb{A10}
\eea

The free ${\cal N}=2$ gauge theory superfield action is
\be
S^{(2)}({\cal W},\bar
{\cal W})=\frac1{4f^2}\int d^8{\cal Z} {\cal W}^2 +\frac1{4f^2}\int
d^8\bar{\cal Z}\bar{\cal W}^2\,, \lb{free}
\ee
where  $f$ is a coupling constant of the dimension $-2$.
Respectively, we  ascribe to ${\cal W}$ the non-standard dimension, $ [{\cal W}]=-1\,$, in accordance with the
interpretation of ${\cal W}$ as the Goldstone superfield associated with a central charge of the mass dimension $1$ in
${\cal N}=4, d=4$ superalgebra \cite{BIK00}.
The free equation of motion corresponding to the action \p{free} is
\be
D^{kl} {\cal W}-\bar D^{kl} \bar {\cal W}=0\,.\lb{A12}
\ee

The  nonlinear $R$ invariant superfield interaction can be written as:
\be
S_{int}({\cal W})=\frac1{f^2}\int d^{12}Z\, L({\cal W})\,,\lb{A13}
\ee
where  the  superfield density $L$ has the dimension $-4$ and may depend on
various dimensionful superfield arguments
\be
{\cal W}^2\bar{\cal W}^2,\quad {\cal W}^{2+k}\Box^k\bar{\cal W}^{2+k},\ldots\,,
\ee
as well as the dimensionless $R$  invariant variables
\bea
&& \bar{D}^4\bar {\cal W}^2~,\quad \bar{D}^4[\bar {\cal W}^2(D^4W^2)]~,
D^{kl} {\cal W},\quad \partial^m{\cal W}\partial_m\bar{\cal W},\ldots\,.
\eea
We can rescale the action according to
\be
S_{int}({\cal W})\rightarrow \frac1{l^2}S_{int}(l{\cal W}) \lb{rescale}
\ee
and make use of this freedom to set, for simplicity, $f=1$.
Throughout the paper, we stick just to this choice.

Introducing the variational derivative of the action $S=S_2+S_{int}$
with respect to the chiral superfield strength,
\bea
&&{\cal M}\equiv
-2i\frac{\delta S}{\delta {\cal W}},\quad \bar{\cal M}\equiv
2i\frac{\delta S}{\delta\bar {\cal W}}\,,\lb{A15}
\eea
we can write the nonlinear equation of motion corresponding to the sum of the actions \p{free} and \p{A13} as
\be
D^{kl}{\cal M}-\bar D^{kl} \bar{\cal M}=0\,.\lb{fullEq}
\ee

The nonlinear $O(2)$ duality transformation  mixing the
equation of motion \p{fullEq}
with the Bianchi identity \p{A9} reads:
\be
\delta_\omega {\cal W}=\omega\, {\cal M}({\cal W},\bar {\cal W})\,,\quad
\delta_\omega {\cal M}=-\omega\, {\cal W}\,,\lb{O2}
\ee
where $\omega$ is a real parameter.
The nonlinear integral $O(2)$ self-duality constraint on the interaction
$L$ was given in \cite{KT}:
\be
\int d^8{\cal Z}({\cal W}^2+{\cal M}^2)= \int
d^8\bar{\cal Z}(\bar {\cal W}^2+\bar{\cal M}^2)\,.\lb{sdint1}
\ee
This condition by itself  is invariant under the duality transformations \p{O2}.

\setcounter{equation}0
\renewcommand{\theequation}{B.\arabic{equation}}
\section*{Appendix B. The recursion calculation of  $L^{(12)}$ }
The expression for the 12th order term $L^{(12)}$ follows from
the general formula for the Lagrange density \p{B7bb}
\bea
&&L^{(12)}=\frac14{\cal A}_0^{(10)}\bar{\cal A}_0^{(2)}
+\frac14{\cal A}_0^{(8)}\bar{\cal A}_0^{(4)}
+\frac14{\cal A}_0^{(6)}\bar{\cal A}_0^{(6)}
+\frac14{\cal A}_0^{(4)}\bar{\cal A}_0^{(8)}
+\frac14{\cal A}_0^{(2)}\bar{\cal A}_0^{(10)}
\nn
&&-\frac1{16}{\cal A}_1^{(9)}\Box\bar{\cal A}_1^{(3)}
-\frac1{16}{\cal A}_1^{(7)}\Box\bar{\cal A}_1^{(5)}
-\frac1{16}{\cal A}_1^{(5)}\Box\bar{\cal A}_1^{(7)}
-\frac1{16}{\cal A}_1^{(3)}\Box\bar{\cal A}_1^{(9)}
\nn
&&
+\frac1{64}{\cal A}_2^{(8)}\Box^2\bar{\cal A}_2^{(4)}
+\frac1{64}{\cal A}_2^{(6)}\Box^2\bar{\cal A}_2^{(6)}
+\frac1{64}{\cal A}_2^{(4)}\Box^2\bar{\cal A}_2^{(8)}
\nn
&&
-\frac1{256}{\cal A}_3^{(7)}\Box^3\bar{\cal A}_3^{(5)}
-\frac1{256}{\cal A}_3^{(5)}\Box^3\bar{\cal A}_3^{(7)}
+\frac1{1024}{\cal A}_4^{(6)}\Box^4\bar{\cal A}_4^{(6)}.
\eea
Recall that it is real up to a total derivative. Below we outline the necessary steps in calculation of this superfield.

The functions ${\cal A}_3^{(7)}$ and ${\cal A}_4^{(6)}$ are known
from eqs. \p{Areg}:
\bea
{\cal A}_3^{(7)}=\frac4{15}{\cal W}^5 \bar{D}^4\bar{\cal W}^2\,,\quad {\cal A}_4^{(6)}=\frac2{45}{\cal W}^6\,.
\eea
The unknown functions ${\cal A}_1^{(9)}$ and ${\cal A}_2^{(8)}$ can be found from the recursion equations
\bea
&& \partial{\cal A}_2^{(8)}=2{\cal A}_1^{(7)}
-Z^{(3)}{\cal A}_2^{(4)}-Z^{(1)}{\cal A}_2^{(6)},\nn
&&\partial{\cal A}_1^{(9)}=2{\cal A}_0^{(8)}-
Z^{(5)}{\cal A}_1^{(3)}-Z^{(3)}{\cal A}_1^{(5)}-Z^{(1)}{\cal A}_1^{(7)}\,.\lb{Zhigh}
\eea
One can directly integrate these equations, using, e.g., the primitives
\bea
&&~ \int_{\cal W}{\cal W}^2\bar{\cal W}^2(D^4{\cal W}^2)^2\nn
&&
=\frac13{\cal W}^3\bar{\cal W}^2(D^4{\cal W}^2)^2
+\frac13{\cal W}^4\bar{\cal W}^2 (D^4{\cal W}^2)\,\Box \bar{\cal W}
+\frac2{15}{\cal W}^5\bar{\cal W}^2(D^4{\cal W})^2,\nn
&&\int_{\cal W} {\cal W}^2(\bar{\cal W}^2) (D^4{\cal W}^2) \bar{D}^4\bar{\cal W}^2
\nn
&&
=\frac13{\cal W}^3\bar{\cal W}^2 (D^4{\cal W}^2) (\bar{D}^4\bar{\cal W}^2)
+\frac16{\cal W}^4\bar{\cal W}^2 (\bar{D}^4\bar{\cal W}^2) \Box \bar{\cal W}\, .\lb{APint}
\eea
It is also straightforward to calculate other primitives needed for solving eqs. \p{Zhigh}. As the result,
we obtain
\bea
{\cal A}_2^{(8)}
= {\cal W}^4\bar{D}^4\Big\{\frac12\bar{\cal W}^2 \bar{D}^4\bar{\cal W}^2
+\frac14\bar{\cal W}^2 D^4{\cal W}^2- \frac1{15}{\cal W}\Box\bar{\cal W}^3\Big\},
\eea
\bea
&& {\cal A}_1^{(9)}=\bar{D}^4\Big\{\frac16{\cal W}^3\bar{\cal W}^2 [(D^4{\cal W}^2)^2 + 3 (D^4{\cal W}^2) (\bar D^4\bar{\cal W}^2)
+ 2(\bar D^4\bar{\cal W}^2)^2] \nn
&& + \, \frac16 {\cal W}^4 \bar{\cal W}^2 (D^4{\cal W}^2) \Box\bar{\cal W} - \frac1{12} {\cal W}^4\bar{\cal W}(D^4{\cal W}^2)\Box \bar{\cal W}^2 +
\frac{1}{4}{\cal W}^4 \bar{\cal W}^2(\bar D^4\bar {\cal W}^2) \Box \bar{\cal W} \nn
&&-\, \frac{1}{4}{\cal W}^4 \bar{\cal W}(\bar D^4\bar {\cal W}^2) \Box \bar{\cal W}^2
- \frac1{12}{\cal W}^4 (\bar D^4\bar {\cal W}^2) \Box \bar{\cal W}^3 + \frac16 {\cal W}^3 \bar{\cal W}^2 D^4({\cal W}^2 \bar D^4\bar{\cal W}^2)\nn
&& - \,\frac1{27} {\cal W}^3\bar{\cal W}^3\Box D^4{\cal W}^3
+ \frac1{36} {\cal W}^4\bar{\cal W}^3\Box D^4{\cal W}^2 -\frac1{12} {\cal W}^4\bar{\cal W} \Box(\bar{\cal W}^2D^4{\cal W}^2) \nn
&& -\,\frac1{36}{\cal W}^4\Box(\bar{\cal W}^3 D^4{\cal W}^2) + \frac1{15}{\cal W}^5\bar{\cal W}^2 (\Box\bar{\cal W})^2 +
\frac1{90}{\cal W}^5\bar{\cal W}^3\Box^2\bar{\cal W} \nn
&& -\,\frac1{30}{\cal W}^5\bar{\cal W}(\Box\bar{\cal W}^2)\Box\bar{\cal W} -\frac1{30}{\cal W}^5\bar{\cal W}\Box(\bar{\cal W}^2\Box\bar{\cal W})
+ \frac1{40}{\cal W}^5\Box(\bar{\cal W}^2\Box\bar{\cal W}^2)\nn
&& -\,\frac1{90}{\cal W}^5\Box(\bar{\cal W}^3\Box\bar{\cal W}) + \frac1{720}{\cal W}^5 \Box^2\bar{\cal W}^4\Big\}.
\eea
The correctness of these expressions was confirmed by checking that they satisfy the $\bar Z$ counterparts of eqs. \p{Zhigh},
\bea
&& \bar\partial{\cal A}_2^{(8)}= \frac12 \Box{\cal A}_3^{(7)}
-\bar Z^{(3)}{\cal A}_2^{(4)}- \bar Z^{(1)}{\cal A}_2^{(6)},\nn
&&\bar\partial{\cal A}_1^{(9)}= \frac12 \Box{\cal A}_2^{(8)}-
\bar Z^{(5)}{\cal A}_1^{(3)}- \bar Z^{(3)}{\cal A}_1^{(5)}- \bar Z^{(1)}{\cal A}_1^{(7)}\,.\lb{barZhigh}
\eea

The explicit expressions for $L^{(12)}$ and $I_{BI}^{(12)}$ can now be directly written, but they are too bulky to present them here.
Note that $L^{(12)}$ could be equivalently calculated by the method of section 2.3.

\setcounter{equation}0
\renewcommand{\theequation}{C.\arabic{equation}}
\section*{Appendix C. The bricks of the action $I^{(10)}_{BI}$}

In this appendix we calculate explicitly different terms in the 10th order
action
$\frac12\int d^8{\cal Z}\,{\cal A}^{(10)}_0=\int d^{12}Z\,L^{(10)}\,$,
where ${\cal A}^{(10)}_0$ is the chiral density \p{A100}.
\bea
&&\frac14\int d^{12}Z\bar{\cal W}^2{\cal A}_0^{(8)}
+\frac14\int d^{12}Z{\cal W}^2\bar{\cal A}_0^{(8)}\nn
&&=\frac1{8}\int d^{12}Z\Big\{\frac1{4}{\cal W}^2\bar{\cal W}^2(D^4{\cal W}^2)^3
+\frac3{4}{\cal W}^2\bar{\cal W}^2(D^4{\cal W}^2)^2(\bar{D}^4\bar{\cal W}^2)
+\frac3{4}{\cal W}^2\bar{\cal W}^2(D^4{\cal W}^2)(\bar{D}^4\bar{\cal W}^2)^2
\nn
&&+\frac1{4}{\cal W}^2\bar{\cal W}^2(\bar{D}^4\bar{\cal W}^2)^3
+\frac1{4}{\cal W}^2\bar{\cal W}^2(\bar{D}^4\bar{\cal W}^2)D^4[{\cal W}^2
(\bar{D}^4\bar{\cal W}^2)]
+\frac1{4}{\cal W}^2\bar{\cal W}^2(D^4{\cal W}^2)\bar{D}^4[\bar{\cal W}^2
(D^4{\cal W}^2)]\nn
&&-\frac1{18}{\cal W}^2\bar{\cal W}^3(\bar{D}^4\bar{\cal W}^2)\Box D^4{\cal W}^3
-\frac1{18}{\cal W}^3\bar{\cal W}^2(D^4{\cal W}^2)\Box \bar{D}^4\bar{\cal W}^3
\nn
&&-\frac1{6}{\cal W}^3\bar{\cal W}^2(\bar{D}^4\bar{\cal W}^2)
\Box\bar{D}^4\bar{\cal W}^3
-\frac1{6}{\cal W}^2\bar{\cal W}^3(D^4{\cal W}^2)\Box D^4{\cal W}^3\nn
&&-\frac2{9}{\cal W}^3(\bar{D}^4\bar{\cal W}^2)\Box[(\bar{\cal W}^3(D^4{\cal W}^2)]
+\frac1{144}{\cal W}^4(\bar{D}^4\bar{\cal W}^2)\Box^2\bar{\cal W}^4
+\frac1{144}\bar{\cal W}^4(D^4{\cal W}^2)\Box^2{\cal W}^4\Big\}.\lb{C1}
\eea

\bea
&&\frac14\int d^{12}Z[\bar{\cal A}_0^{(4)}{\cal A}_0^{(6)}
+{\cal A}_0^{(4)}\bar{\cal A}_0^{(6)}]\nn
&&=\frac1{8}\int d^{12}Z\left\{\frac1{4}{\cal W}^2\bar{\cal W}^2
[(D^4{\cal W}^2)+(\bar{D}^4\bar{\cal W}^2)]
\bar{D}^4[\bar{\cal W}^2(D^4{\cal W}^2)]\right.\nn
&&+\frac1{4}{\cal W}^2\bar{\cal W}^2
[(D^4{\cal W}^2)+(\bar{D}^4\bar{\cal W}^2)]D^4[{\cal W}^2
(\bar{D}^4\bar{\cal W}^2)]\nn
&&\left.-\frac1{18}\bar{D}^4[\bar{\cal W}^2(D^4{\cal W}^2)]
({\cal W}^3\Box\bar{\cal W}^3)-\frac1{18}D^4[{\cal W}^2(\bar{D}^4\bar{\cal W}^2)]
(\bar{\cal W}^3\Box {\cal W}^3)\right\}.\lb{C2}
\eea

\bea
&&-\frac1{16}\int d^{12}Z[{\cal A}_1^{(7)}\Box \bar{\cal A}_1^{(3)}
+\bar{A}_1^{(7)}\Box {\cal A}_1^{(3)}]\nn
&&=\frac1{8}\int d^{12}Z\left\{
-\frac1{6}{\cal W}^3(\bar{D}^4\bar{\cal W}^2)^2\Box\bar{\cal W}^3
-\frac1{9}{\cal W}^3\bar{D}^4[\bar{\cal W}^2(D^4{\cal W}^2)]
\Box\bar{\cal W}^3\right.\nn
&&-\frac1{6}\bar{\cal W}^3{\cal W}^2(D^4{\cal W}^2)\Box D^4{\cal W}^3
-\frac1{9}\bar{\cal W}^3{\cal W}^2(\bar{D}^4\bar{\cal W}^2)\Box D^4{\cal W}^3\nn
&&\left.
+\frac1{36}{\cal W}^4(\Box\bar{D}^4\bar{\cal W}^3)\Box\bar{\cal W}^3
+\frac1{36}\bar{\cal W}^4\Box {\cal W}^3(\Box D^4{\cal W}^3)\right\}.\lb{C3}
\eea
\bea
&&-\frac1{16}\int d^{12}Z{\cal A}_1^{(5)}\Box \bar{\cal A}_1^{(5)}=
-\frac1{36}\int d^{12}Z {\cal W}^3(\bar{D}^4 \bar{\cal W}^2)
\Box[\bar{\cal W}^3(D^4 {\cal W}^2)].\lb{C4}
\eea
\bea
&&\frac1{64}\int d^{12}Z[{\cal A}_2^{(6)}\Box^2 \bar{\cal A}_2^{(4)}
+\bar{\cal A}_2^{(6)}\Box^2 {\cal A}_2^{(4)}]
\nn
&&=
\frac1{384}\int d^{12}Z[{\cal W}^4(\bar{D}^4\bar{\cal W}^2)
\Box^2(\bar{\cal W}^4)+\bar{\cal W}^4(D^4{\cal W}^2)
\Box^2({\cal W}^4)].\lb{C5}
\eea
\bea
&&
-\frac1{256}\int d^{12}Z{\cal A}_3^{(5)}\Box^3\bar{\cal A}_3^{(5)}
=-\frac1{14400}
\int d^{12}Z{\cal W}^5\Box^3\bar{\cal W}^5.\lb{C6}
 \eea

Summing up the terms \p{C1} - \p{C6} yields just the action \p{10ord}.

\end{document}